\documentclass[superscriptaddress, nofootinbib, preprint]{revtex4}[12pt]
\usepackage{graphicx}
\usepackage {amsmath, amssymb, rotating}
\usepackage{subcaption}
\usepackage{braket}
\usepackage{footnote}
\usepackage[usenames,dvipsnames]{color}
\usepackage{mdwlist} 
\usepackage{slashed}
\usepackage{verbatim}
\usepackage{mathrsfs}

\newcommand{\be}{\begin{eqnarray}}
\newcommand{\ee}{\end{eqnarray}}

\newcommand{\CO}{\mathcal{O}}
\newcommand{\beq}{\begin{eqnarray}}
\newcommand{\eeq}{\end{eqnarray}}

\begin{document}

\title{Detecting Dark Blobs}%

 \author{Dorota M Grabowska}%
\email{dgrabowska@berkeley.edu}
\affiliation{Department of Physics, University of California, Berkeley, California 94720, USA}
\affiliation{Theoretical Physics Group, Lawrence Berkeley National Laboratory, Berkeley, California 94720, USA}

 \author{Tom Melia}%
\email{tom.melia@ipmu.jp}
\affiliation{Kavli Institute for Physics and Mathematics of the Universe (WPI), University of Tokyo, Kashiwa, 277-8583, Japan}

 \author{Surjeet Rajendran}%
\email{surjeet@berkeley.edu}
\affiliation{Department of Physics, University of California, Berkeley, California 94720, USA}

\preprint{IPMU18-0113}%

\begin{abstract}{
Current dark matter detection strategies are based on the assumption that the dark matter is a gas of non-interacting particles with a reasonably large number density. This picture is dramatically altered if there are significant self interactions within the dark sector, potentially resulting in the coalescence of dark matter particles into large composite blobs. The low number density of these blobs necessitates new detector strategies. We study cosmological, astrophysical and direct detection bounds on this scenario and identify experimentally accessible parameter space. The enhanced interaction between large composite states and the standard model allows searches for such composite blobs using existing experimental techniques. This includes the detection of scintillation in MACRO,  XENON and LUX, heat in calorimeters such as CDMS, acceleration and strain in gravitational wave detectors such as LIGO and AGIS, and spin precession in CASPEr. These searches leverage the fact that the transit of the dark matter occurs at a speed $\sim$ 220 km/s, well separated from relativistic and terrestrial sources of noise. They can be searched for either through modifications to the data analysis protocol or relatively straightforward adjustments to the operating conditions of these experiments. 
}
\end{abstract}

\maketitle
\section{Introduction}
\label{sec:intro}
Identifying the nature of dark matter is one of the great open challenges in physics. Discovery of the non-gravitational properties of dark matter would provide a portal into a new sector of particle physics and may shed light on its unique role in structure formation. All  current dark matter detection strategies, ranging from direct detection efforts in the laboratory to indirect signals from the annihilation (or decay) of dark matter, are based on the assumption that the dark matter is distributed around the universe as a gas of free particles with a reasonably large number density.\footnote{The exception are searches for dark matter with astrophysical scale mass, such as Primordial Black Holes.} This large number density yields a high enough flux of dark matter enabling the detection of rare dark matter events. 

This picture of dark matter as a gas of free particles naturally emerges if self interactions within the dark sector are weak. What if the dark sector had strong self interactions? 
In this case, much like the standard model undergoing nucleosynthesis and producing  composite nuclei, the dark sector will also undergo a nucleosynthesis process in the Early Universe that may be highly efficient since it need not suffer from the accidents of nuclear physics in the standard model that inhibit the production of heavy elements. As a result, individual dark matter particles could coalesce to form very large composite states   \cite{Wise:2014ola, Hardy:2014mqa, Gresham:2017cvl}\footnote{Note that this requires a particle - antiparticle asymmetry in the dark sector, \cite{Wise:2014ola, Hardy:2014mqa}.}. (See also Refs.~\cite{Krnjaic:2014xza,Detmold:2014qqa} for further examples of dark matter nucleosynthesis resulting in more modestly sized states.) Observational constraints on these self-interactions are weak.  The most stringent constraints arise from observations of the Bullet Cluster, restricting these self interaction cross-sections to be less than approximately $1 \text{ cm}^2/\text{gm}$ \cite{Spergel:1999mh, Kahlhoefer:2013dca}. Since this bound is based on the dark matter distribution today, it is significantly weakened if the dark matter is clustered into heavy composite states with a low number density.

In this paper, we study observational limits on such large composite states of dark matter and propose generic experimental strategies that could be employed to search for them in the laboratory. We will henceforth refer to these states as dark blobs. Our investigation is restricted to dark blobs with mass less than $10^{33}$ GeV, so that at least one dark blob passes through the Earth in a year, enabling the possibility of direct detection. The challenge in detecting this type of dark matter arises from the fact that their number density is low, necessitating detectors of large volume. However, the large number of constituents in the blob enhances the scattering cross-section between the blob and a detector - in particular, significant enhancements are possible if the scattering is coherent---see Ref.~\cite{Hardy:2015boa} for a detailed study of direct detection form factors. This potentially enables multiple observable interactions of the dark blob with the standard model.  Our detection strategy will leverage the fact that the transit of dark matter occurs at a speed $v \sim 10^{-3}$, characteristic of dark matter. This speed lies in an interesting window between terrestrial sources of noise and the relativistic speeds  of cosmic ray events. Moreover, events induced by the dark matter should also lie along a straight line, enabling an additional background discriminant. 

We model the blob as a composite state of dark matter, consisting of a large number of dark matter particles bound by some self-interaction in the dark sector. We assume that the total mass of the blob that contains $N_X$ constituents, hereby labelled as $\chi$, of mass $m_\chi$ is $M_X = N_X\, m_\chi$. The Bohr radius of the constituents in the bound state is $\Lambda_{\chi}^{-1}$. The scale $\Lambda_{\chi}$ (through a form factor) determines the momenta that can be exchanged between the blob and the standard model. The binding energy of $\chi$ to the blob is also a function of $\Lambda_{\chi}$, though this energy will not play as direct a role in our phenomenology (see e.g. Refs.~\cite{Wise:2014jva,Wise:2014ola,Gresham:2017zqi} for a study of large composite state structure). 

The primary objective of this paper is to explore the qualitative features of the phenomenology of these blobs and to establish the robustness of the parameter space that could be experimentally accessed. We establish these by studying a restricted range of parameters, where these qualitative features can be unambiguously seen. Therefore, in this paper, we limit our investigation to interactions between blobs and nuclei, though the mediator may directly couple to either nucleons or photons. Moreover, we will mostly only study the parameter space where  $10 \text{ keV } \lessapprox \Lambda_\chi \lessapprox 10$ MeV. This is because the maximum momentum that can be exchanged between $\chi$ and a probe  is the smaller of $\Lambda_\chi$ and the momentum of the probe.\footnote{This arises from bound state form-factors, see Appendix~\ref{sec:BNscattering} and Ref.~\cite{Hardy:2015boa}.} In a terrestrial experiment, a probe nucleus will collide with the blob with a momentum $\sim$ 10 MeV.  The phenomenology of the blobs when $\Lambda_\chi \gtrapprox$ 10 MeV should thus be similar to that of the case where $\Lambda_\chi \sim 10$ MeV and will thus not be separately analyzed. The lower limit  $\Lambda_\chi \sim$ 10 keV  is imposed for convenience. As we will see below, this limit makes it easier to treat the coherent scattering of a bosonic blob in high density matter. Moreover, it also enables us to ignore inelastic excitations of the blob due to its interactions with the standard model. Finally, for simplicity, in this paper we only consider the case where $\Lambda_\chi \approxeq m_{\chi}$. 

The phenomenology of the blob changes drastically depending upon whether the constituents $\chi$ are bosonic or fermionic. The size of a bosonic bound state is independent of the number of constituents in that state, while Pauli exclusion forces the fermionic blob to increase its size as it grows in mass. We thus study the bosonic and fermionic cases separately, in sections \ref{sec:bosons} and \ref{sec:fermions} respectively. In each section, we begin our analysis by computing the scattering cross-section between the blob and the standard model. This cross-section is used to compute observational bounds on the blob from terrestrial, astrophysical and cosmological observations. We assume that the constituents $\chi$ of the blob and the standard model interact with each other through a mediator $\phi$ of mass $\mu$. When the mediator has a range shorter than the de-Broglie wavelength of the standard model probe, the interaction has to be described using quantum mechanics. Long range mediators can be treated classically. 
 
 Irrespective of the microphysics of the blobs, there are only four possible experimental signatures of the interaction of the dark blob with nucleons. For short range mediators, the only possible effect is the deposition of energy by the blob when it collides with nucleons.  This energy may be sufficient to ionize the standard model probe  or may simply dump energy into acoustic modes. Energy can also be similarly deposited if the blob exerts long range forces on nuclei. In addition to energy deposition, a classical field can have three other physical effects: it can induce precession of spins, accelerate matter and change the values of fundamental constants. Of these experimental signatures, ionization is presently searched for in a number of experiments and is well constrained. We propose new experimental techniques to search for the other effects in section \ref{sec:detection}.  

\section{Bosonic Blob}
\label{sec:bosons}
We consider a blob of mass $M_X$ consisting of $N_X = M_X/m_{\chi}$ particles, each of mass $m_{\chi}$. This blob is spread over a distance $\Lambda_\chi^{-1}$. A model independent bound can be placed on $N_X$ by demanding that the mass confined within $\Lambda_\chi$ does not form a black hole - this mass is $\sim M_{Pl}^2/\Lambda_\chi \gg 10^{33} \text{ GeV}$, the largest mass that is of interest in a terrestrial detector, for $\Lambda_\chi \lessapprox 10 \text { MeV}$. After computing the scattering cross-section between the blob and the standard model in sub-sections \ref{sec:bosonshortrange} and \ref{sec:bosonlongrange}, we discuss the bounds on the mediator and evaluate observational constraints on this kind of blob in sub-section \ref{sec:bosonconstraints}. 

\subsection{Short Range}
\label{sec:bosonshortrange}
We assume that $\chi$ interacts with the standard model through a scalar mediator $\phi$ through the Lagrangian: 
\be
\mathcal{L} \supset g_{\chi} m_{\chi} \phi \chi^{*} \chi + g_{N} \phi \bar{\Psi}_{N} \Psi_{N}
\label{eq:bosshort}
\ee
where $\Psi_{N}$ represents a nucleon. Upon integrating out $\phi$, the effective interaction is described by a contact operator, $\CO_c$,  
\beq
\CO_c \sim \frac{g_{\chi} g_{N} m_{\chi} \chi^{*}\chi \bar{\Psi}_{N}\Psi_{N}}{\mu^2}
\label{eq:BosShortRangeOperator}
\eeq
where $\mu$ is the mass of the mediator. The coupling between the dark matter and the mediator induces additional self-interactions in the blob, which limits the number of constituents in a stable blob. Specifically, this coupling leads to a quartic self-interaction term $g_{\chi}^2 m_{\chi}^2/\mu^2$ between the constituents. In order for the blob to be stable, the energy induced by this quartic term must be smaller that the energy arising from the quadratic term, ie
\beq
\left(g_{\chi}^2 m_{\chi}^2/\mu^2\right) \chi_{s}^4 \, \lesssim \, m_{\chi}^2 \chi_{s}^2 \qquad \qquad \chi_s \sim  \sqrt{\left(M_{X}/m_{\chi}^2\right) \Lambda_\chi^3} 
\eeq
where $\chi_s$ is the classical field value of $\chi$ in the blob. Assuming $m_{\chi} \sim \Lambda_\chi$, we find that $g_\chi$ is constrained to be
\begin{eqnarray}
g_{\chi} \lessapprox \left(\mu/\Lambda_\chi\right) N_{X}^{-1/2}\,.
\label{eq:bosstab}
\end{eqnarray}
Note that there may be ways to model-build around this constraint by introducing additional interactions, though we do not pursue this in this paper.

There are two important effects that determine the scattering cross section between a blob and nucleus. First, the scattering occurs between the constituents ($\chi$) in the blob and the nucleon and therefore the momentum $q$ that can be transferred in such a process is determined by the bound state wave-function of the constituent; such form factors have been studied in the context of blobs in \cite{Hardy:2015boa}. This inhibits momentum transfers $q \gg \Lambda_\chi$. Second, the scattering cross-section between the blob and the nucleon can be coherently enhanced by the number of constituents in the blob. The details of this calculation are summarized in Appendix \ref{sec:BNscattering}. Using this result, the differential scattering cross-section off a nucleon is
\be
\frac{d\sigma}{d\Omega} =  N_X^2 \left(\frac{g_{\chi}^2 g_{N}^2 m_N^2}{\mu^4}\right) F_{B}\left(q/\Lambda_\chi\right) 
\ee
where $F_{B}$ is a form factor that suppresses momentum transfers between the blob and the nucleons that are bigger than the Bohr-radius $\Lambda_\chi^{-1}$. This scattering cross-section is coherently enhanced by $N_X^2$ as long as the momentum transfer and the de Broglie wavelength of the probe nucleon is larger than $\Lambda_\chi^{-1}$. By choosing $\Lambda_\chi \gtrapprox 10$ keV, this condition is satisfied for a typical terrestrial detector, where one might consider the collision of a nucleus at temperatures $\sim 300$ K with the blob. For a short range mediator, the above coherently enhanced cross-section is cut-off by the geometric size of the blob. Hence, the total scattering cross-section between the blob and a nucleon is
\be
\sigma \simeq \text{Min}\left(N_X^2 \frac{g_{\chi}^2 g_N^2 m_N^2}{\mu^4}  \frac{\Lambda_\chi^2}{m_N^2 v_{\chi}^2}, \frac{1}{\Lambda_\chi^2}\right)
\label{eq:shortbosxsec}
\ee
where the suppression factor $\Lambda_{\chi}^2/\left(m_N^2 v_{\chi}^2\right)$ arises from the fact that the form factor suppresses momentum transfer larger than $\Lambda_{\chi}$, leading to a reduction in the phase space available for scattering.

\begin{figure} 
\includegraphics[width=.9\textwidth]{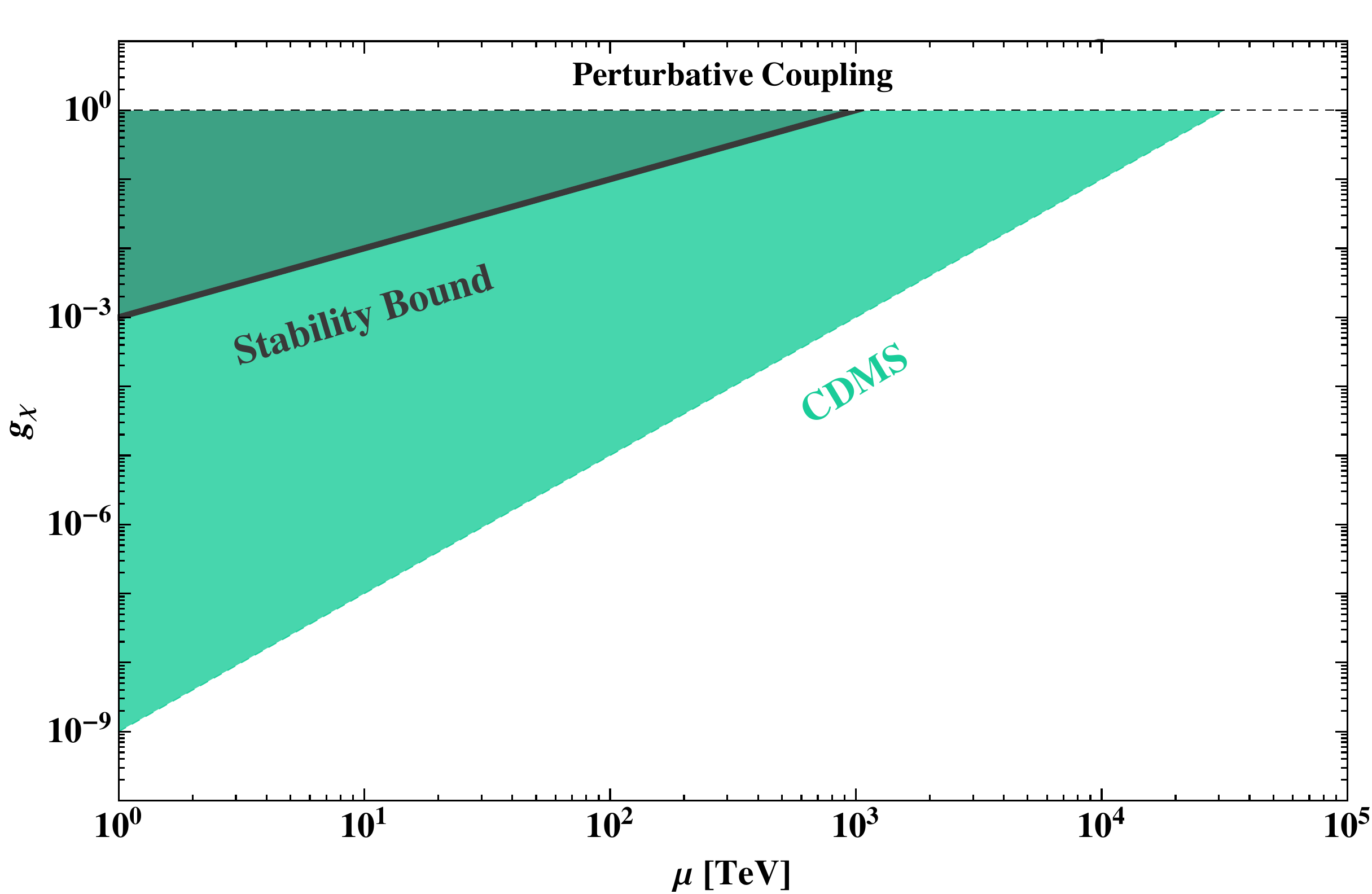}
\caption{Reach on coupling $g_\chi$ for a bosonic blob of mass $M_X=10^{16}\,\,$GeV and Bohr radius $\Lambda_\chi=100\,$keV, as a function of short-range mediator mass $\mu$. For this value of $\Lambda_\chi$, energy deposition is due to many soft scatterings that are insufficient to cause ionization, yet still be detected in CDMS. The lower bound on $g_\chi$ arises from our requirement that the cross section saturate the geometric cross section bound, {\it i.e.} $\sigma \sim 1/\Lambda_\chi^{2}$.}
\label{fig:BosonShortRange}
\end{figure}

In this paper, we limit our investigation to the parts of parameter space  where  $\sigma = \Lambda_\chi^{-2}$, with the maximum momentum transferred set by $q \sim \Lambda_\chi$. This parameter space is shown in Fig.~\ref{fig:BosonShortRange}. In this range, when a blob transits a medium of  number density $\eta_m$, the energy deposited per unit length in this medium is: 
\be
\frac{dE}{dx} \sim \frac{\Lambda_\chi^2}{m_p} \, \frac{1}{\Lambda_\chi^2} \eta_m \sim \frac{\eta_m}{m_p} \sim \text{keV/cm} \left(\frac{\eta_m}{10^{22}/\text{cm}^3} \right)
\label{eqn:bosededx}
\ee
where $m_p$ is the mass of the probe. In this limit, when the de-Broglie wavelength of the probe is larger than the geometric size of the blob, the total energy deposited by the blob is independent of the physics of the blob. However, the form in which the energy is deposited depends critically on $\Lambda_\chi$. For $\Lambda_\chi \gtrapprox $ MeV, the energy  deposited in an individual collision is large enough to cause ionization. These signals can be searched for in conventional WIMP detection experiments, particularly in low threshold detectors. But, when $\Lambda_\chi \lessapprox $ MeV, the energy deposition occurs through a number of soft scatterings, none of which are sufficient to cause ionization. These soft scatters require a qualitatively different class of dark matter detectors, some of which we discuss in Section \ref{sec:detection}.

When the de-Broglie wavelength $\lambda_p$ of the probe is smaller than the geometric size $\sim \Lambda_\chi^{-1}$ of the blob, as discussed in  Appendix \ref{sec:BNscattering}, coherent enhancements to the cross-section are possible, but the enhancement is limited by the de Broglie wavelength $\lambda_p$ of the probe. This leads to suppressed energy deposition in the medium and we do not consider this case for bosonic constituents.

\subsection{Long Range}
\label{sec:bosonlongrange}
A long range field sourced by the blob can directly cause accelerations, induce spin precessions and change the values of fundamental constants. To analyze these, we consider the Lagrangian: 
\be
\mathcal{L} \supset g_{\chi} m_{\chi} \phi \chi^{*}\chi + \underbrace{ g_{N} \phi \bar{\Psi}_{N} \Psi_{N}}_{\CO_1} + \underbrace{\frac{\partial_{\mu}\phi}{f_{N}}  \bar{\Psi}_{N} \gamma^{\mu}\gamma_5 \Psi_{N}}_{\CO_2} + \underbrace{\frac{\phi}{\alpha_\text{EM} M} F_{\mu\nu}F^{\mu\nu}}_{\CO_3}
\label{eq:boslonglag}
\ee
where the operator $\CO_1$ leads to accelerations, $\CO_2$ causes spin precessions and $\CO_3$ changes the value of $\alpha_\text{EM}$, the fine-structure constant.\footnote{Dilatons can also change electron and nucleon masses; these effects can be similarly analyzed.} As the detection methods for dark blobs rely on the large number of constituents compensating for weak couplings, we require that the interaction between the dark matter and the mediator be spin-independent.

\begin{figure} 
\includegraphics[width=.9\textwidth]{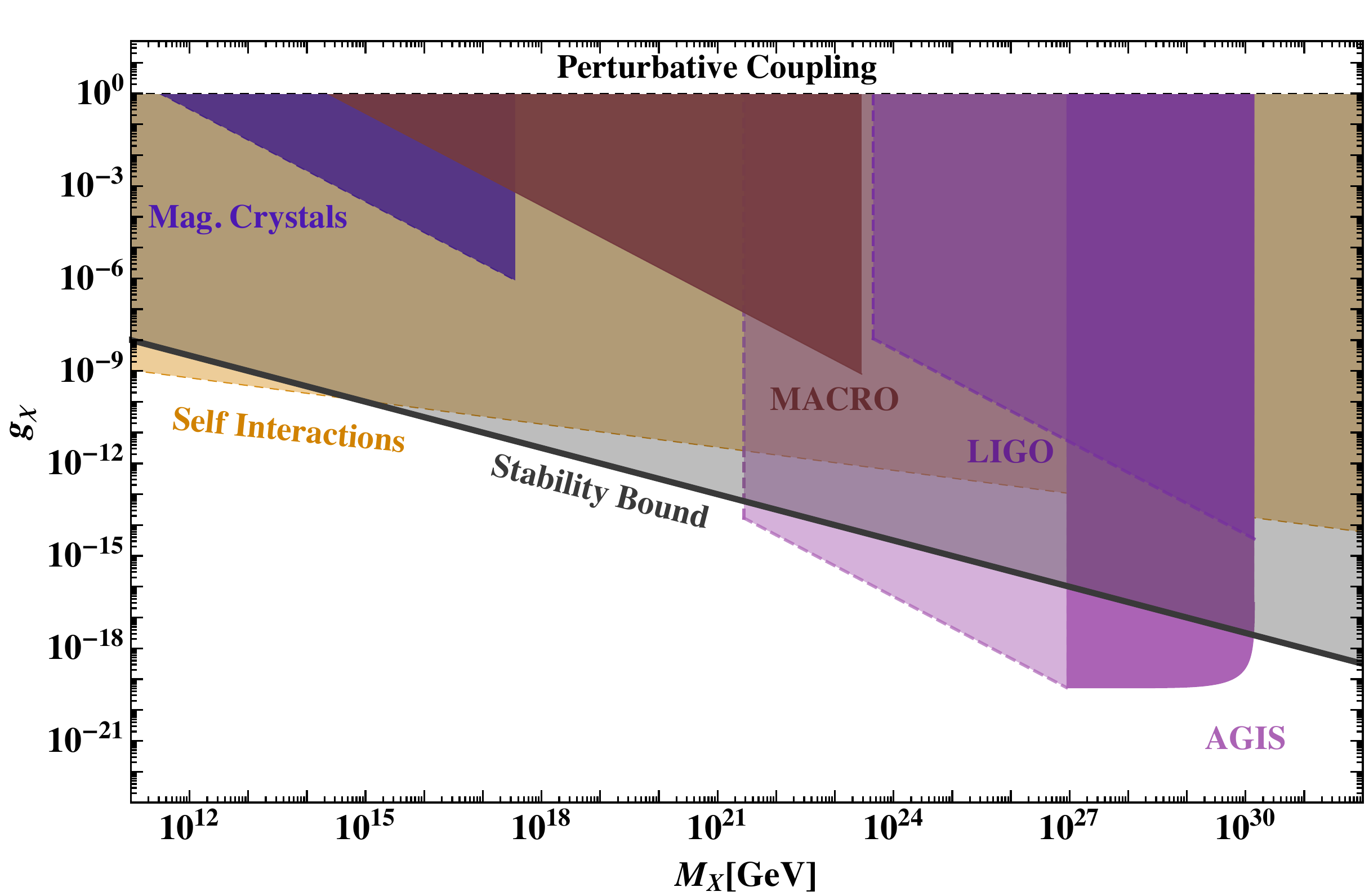}
\caption{Reach on coupling $g_\chi$ for a bosonic blob with Bohr radius $\Lambda_\chi=10\,$keV and a $200\,$km-range scalar mediator, as a function of blob mass $M_X$. Without additional model building, only AGIS is sensitive enough to probe the extremely weak coupling between the mediator and the dark sector. The different shading for this atomic interferometer reach is explained in Sec.~\ref{sec:discussion}.}
\label{fig:AGISBoson}
\end{figure}

\begin{figure} 
\includegraphics[width=.9\textwidth]{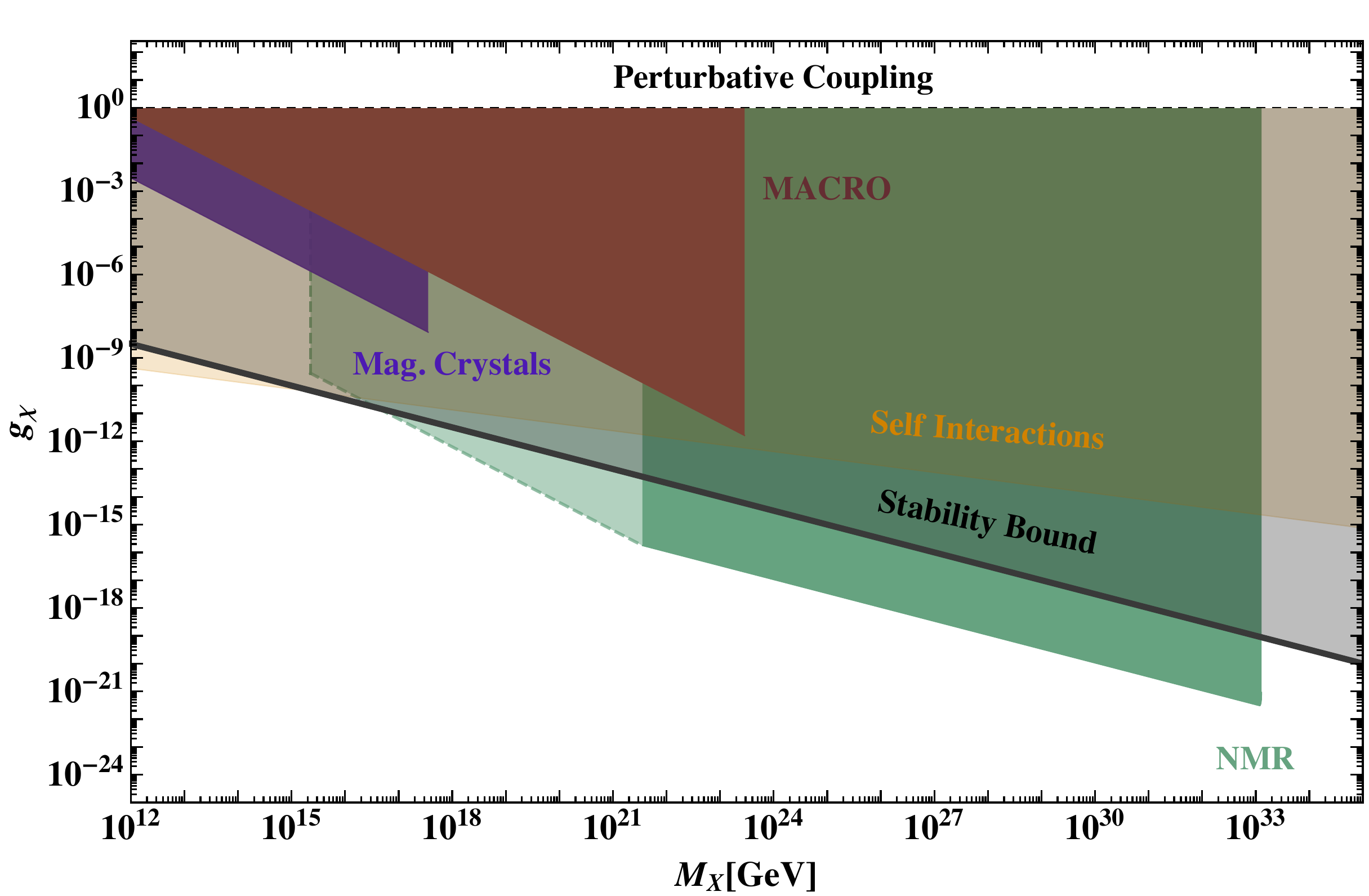}
\caption{Reach on coupling $g_\chi$ for a bosonic blob with Bohr radius $\Lambda_\chi=10\,$keV and a pseudoscalar mediator with a $6000\,$km range, as a function of blob mass $M_X$. Due to the highly compact nature of the bosonic blob, the blob cannot deposit much energy as it passes through detectors that look for either ionization or heat deposition. The different shading for the reach of an NMR-type experiment is explained in Sec.~\ref{sec:discussion}. }
\label{fig:NMRBosons}
\end{figure}

\begin{figure} 
\includegraphics[width=.9\textwidth]{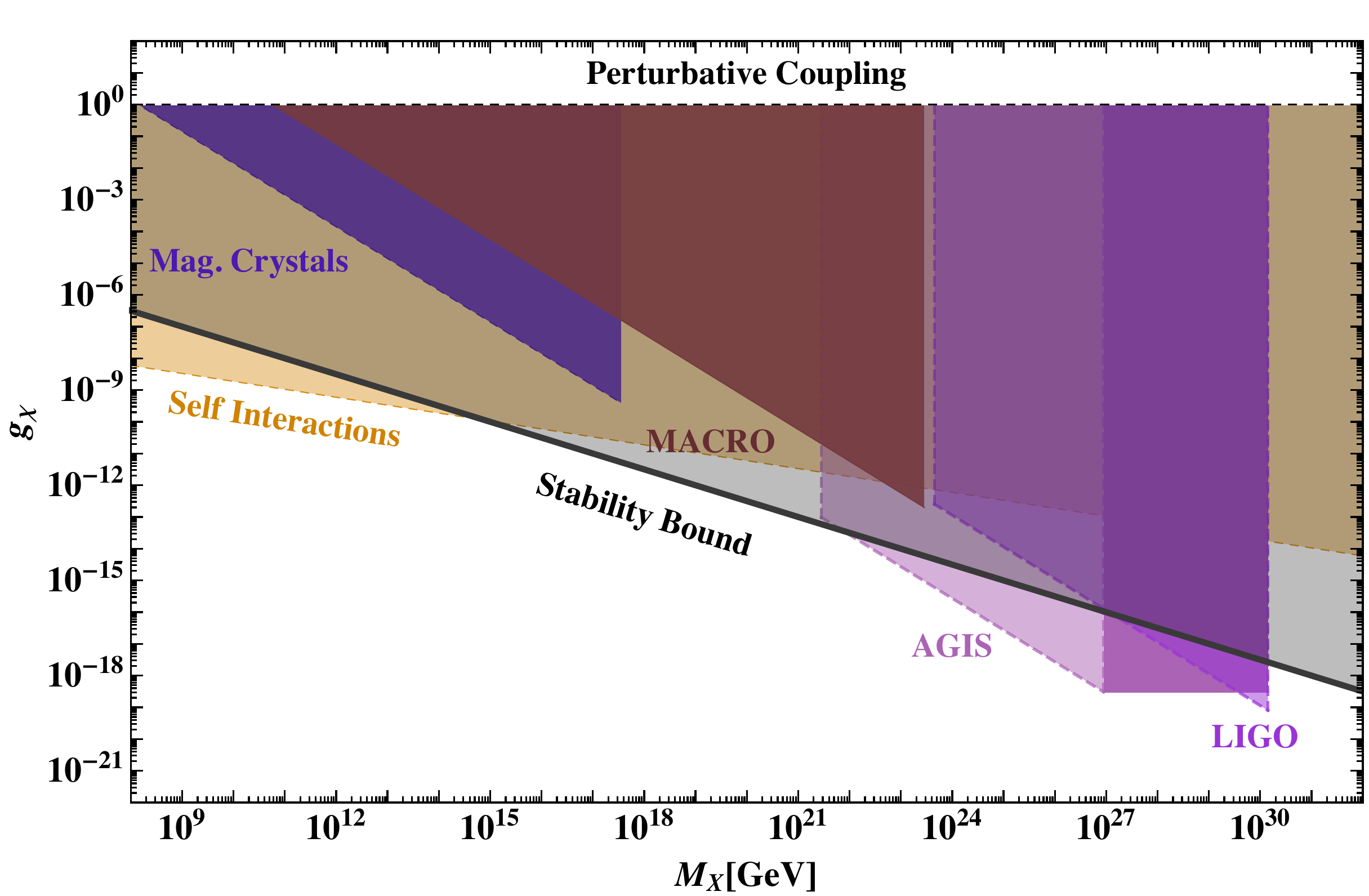}
\caption{Reach on coupling $g_\chi$ for a bosonic blob with Bohr radius $\Lambda_\chi=10\,$keV and a $200\,$km-range dilaton mediator, as a function of blob mass $M_X$. The different shading for the reach of both interferometer experiments is explained in Sec.~\ref{sec:discussion}.}
\label{fig:DilatonBosons}
\end{figure}

Similar to the short range case, the coupling between the constituents and the mediator induces additional self-interactions in the blob itself. When the blob contains a large number of particles, the forces from the classical field sourced by the constituents can destabilize the blob. As a consistency check, we demand that the energy shift caused on a single $\chi$ due to the $N_{\chi}$ particles in the blob is less than $\Lambda_\chi$ {\it i.e.} $N_{\chi} \lessapprox 1/g_{\chi}^2$. In our sensitivity estimates, shown in Figs~\ref{fig:AGISBoson}, \ref{fig:NMRBosons}, and \ref{fig:DilatonBosons} and explained in detail in Sec.~\ref{sec:bosonconstraints} and Sec.~\ref{sec:detection}, we demarcate the regions where the blob satisfies this self consistency check. Note that we also show parameter space that violates this check, as a blob that fails this check might have more complicated dynamics ({\it e.g.} additional stabilizing forces) yet lead to similar observational phenomenology.

As previously mentioned, one way of detecting dark blobs is by looking for the energy deposited by a blob during its transit through a medium. This energy deposited depends not only on the density of the medium $\eta_m$, but also on the speed of the blob and the speed of sound in the medium. In the terrestrial context, the speed of the blob is much larger than the speed of sound in materials and so the energy deposited in this transit can be calculated using the instant approximation (similar to calculations of dynamical friction), leading to energy loss of
\be
\frac{dE}{dx} \sim 2 \pi  \int_{0}^{1/\mu} dr \,  r \,  \eta_m \, M_N \left( \frac{F\left(r\right)}{M_N} \frac{r}{v}\right)^2
\label{eqn:bosedynfric}
\ee
where $F\left(r\right)$ is the force experienced by a probe at a distance $r$ from the blob and $M_N$ is the mass of the medium's constituent nuclei.

When the sound speed in medium is much larger than the speed of dark blob relative to the medium, such as in the Early Universe, Eq.~\ref{eqn:bosedynfric} must be modified. In particular, the relative velocity $v_{C}$ between the blob and the medium is significantly smaller than the speed of sound $c_s$ during baryon acoustic oscillations.  The drag force in this limit can be estimated through the following argument: a baryon of mass $m_p$ at a distance $r$ from the blob experiences a force $F\left(r\right)$. The response time of the medium due to this force is $\sim r/c_s$. The velocity gained by the baryon within this response time is  $\sim \left(\frac{F\left(r\right)}{m_p} \frac{r}{c_s}\right)$. For there to be a drag force on the blob, there must be an asymmetry in the response of the medium to the blob - this arises due to the relative velocity $v_C$ between the blob and the medium. For baryons at a distance $r$ from the blob, the asymmetry due to the motion of the blob in the response time $\sim r/c_s$ is $\sim v_C/c_s$. Thus, the energy deposited is, 
\be
\frac{dE}{dx} \sim 2 \pi \int_{0}^{1/\mu} dr\, r\, \eta_{m}\, m_{p} \left(\frac{F\left(r\right) r}{m_p c_s}\right)^2 \left(\frac{v_C}{c_s}\right) \, .
\label{Eq:BAODrag}
\ee
For a systematic analysis of this drag force, see~\cite{Ostriker:1998fa}.

In addition to energy depositions, the operators $\CO_2$ and $\CO_3$ induce spin precession and strain, yielding qualitatively new experimental signatures. In a terrestrial experiment, the net change in a sample caused by the transit of the blob can also be calculated using the sudden approximation:  the impulse from the transit leads to an instantaneous, potentially observable,  change to the state of the system. For example, consider a spin at a distance $ 1/\Lambda_\chi \lessapprox r \lessapprox 1/\mu$ from the blob {\it i.e.} a spin that is well outside the geometric size of the blob but within the range of $\phi$. This spin rotates by an angle
\be
\delta \theta \sim \frac{g_{\chi} N_X}{f_{N}\,  r \, v}
\label{eq:deltheta}
\ee
due to the transit of the blob. As will be discussed in Sec.~\ref{subsec:spinprecession}, the change in the spin orientation leads to a change in magnetization of a sample that can then be picked up by a SQUID. The operator $\CO_3$ changes the fine-structure constant, inducing strain in materials bound together by electromagnetism. A probe, again at a distance $ 1/\Lambda_\chi \lessapprox r \lessapprox 1/\mu$, experiences a strain 
\be
\label{Eq:Strain}
h \sim \frac{g_{\chi} N_X}{r M}
\label{eq:delh}
\ee
due to the blob. The observable consequences of this strain depends upon the probe, though in general, such strains can be looked for by experiments built to detect gravitational waves. Note that the force induced by $\CO_3$ arises from the electromagnetic contribution to the nuclear mass (of charge $Z$ and atomic mass $A$) and is 
\begin{eqnarray}
F\sim \frac{Z^2}{4\pi}\frac{1}{A^{1/3}\text{fm}} \frac{\alpha_\text{EM}}{M} \nabla \phi \,.
\label{eq:emmassforce}
\end{eqnarray}
We discuss in greater details the methods to search for these novel effects caused by the operators $\CO_2$ and $\CO_3$ in section \ref{sec:detection}. 

\subsection{Constraints}
\label{sec:bosonconstraints}

There are three classes of bounds on this specific dark matter scenario, for both short range and long range mediators. The first class arises from constraints on the mediator $\phi$ due to its interactions with the standard model. The second class arises from direct collisions of the blob with terrestrial experiments. The last class involves astrophysical and cosmological bounds. We discuss these bounds in the following sub-sections. 

\subsubsection{Mediators}
\label{subsubsec:bos5thforce}
The interactions of $\phi$ with the standard model are constrained by a multitude of experiments. As above, we consider two different mediator mass ranges. For long range interactions with the standard model, we consider mediators that have a interaction length scale ranging from a micrometer up to the radius of the Earth: $ \mu\text{m} \lesssim 1/\mu \lesssim 6000 \text{ km}$. In this range, scalar and dilaton couplings are constrained by searches for new forces in the laboratory \cite{Adelberger:2003zx}, while pseudoscalars are limited by astrophysical constraints on stellar cooling, such that $f_N \gtrapprox 10^{10}\text{  GeV}$. For short range interactions, we simply take $g_{N}, g_{\chi} \sim \mathcal{O}\left(1\right)$ with $\mu \gtrapprox$ TeV, whence $\mu$ satisfies  collider constraints. Note that the short range scattering cross-section is geometric ({\it i.e.} $\sigma \sim \Lambda_{\chi}^{-2}$) as long as 
\beq
M_{X} \gtrapprox 10^{8} \text{ GeV } \left(\frac{v_{\chi}}{10^{-3}}\right) \left(\frac{\mu^2}{\text{TeV}^2}\right) \left( \frac{\text{10 keV}}{\Lambda_{\chi}}\right)\, ,
\eeq
easily satisfied in our parameter range.  

The above choice for the mediator range is largely made for simplicity - we wish to demonstrate the experimental viability of this parameter space. A more detailed analysis of the bounds \cite{Green:2017ybv} could reveal additional parts of parameter space where significant scattering might be possible.

\subsubsection{Direct Detection}

The most stringent direct detection constraint on this scenario arises from the MACRO experiment due to its large operating volume~\cite{Ahlen:1992pe, Ambrosio:2002mb, Ambrosio:2002qq}. MACRO is sensitive to ionizing interactions that deposit energies of at least 6 MeV/cm. The energy deposited by short range mediators in our scenario is around keV/cm and these dark matter blobs with bosonic constituents  are not constrained by MACRO. Long range interactions are able to deposit energies of around 10 eV/\AA;  these do cause ionization\footnote{We do not  include the effect of ionization efficiency  for low energy nuclear recoils in making our projections; this can be of order $\lessapprox$ 0.1. \cite{Chavarria:2016xsi}.} and are constrained by MACRO, as shown in Figs~\ref{fig:AGISBoson}, \ref{fig:NMRBosons}, and \ref{fig:DilatonBosons}. For lower energy deposition, the blob will lead to multiple scattering events in direct detection experiments and can be searched for using an optimized search. CDMS has performed searches similar in spirit to this event topology -- the aim was to search for lightly ionizing particles (LIPS)~\cite{Agnese:2014vxh}. However, the LIPS search requires the events to have a profile (in ionization/phonon yield) similar to electron recoils - a restriction that blinds it to blobs depositing energy through nuclear collisions. 

It is important to note that even with an energy deposition of around 10 eV/\AA, for blob masses larger than $\sim 10^{10}$ GeV, the blob will be able to penetrate more than a km of rock overburden and thus  lead to signals in these experiments.  

\subsubsection{Astrophysical and Cosmological Bounds}
\label{subsubsec:boscosmo}

In addition to these direct limits, there are also astrophysical and cosmological bounds on these dark blobs. The blobs have large self-interaction cross-sections that are bounded by observations of merging clusters to be no greater than approximately $1 \text{ cm}^2/\text{gm}$. For bosonic bound states interacting through a short range mediator, this bound is satisfied as long as 
\beq
M_{X} \gtrapprox 10^{-4} \, \text{GeV}^3/\Lambda_{\chi}^2 \, ,
\eeq trivially satisfied in our parameter space. In the case of a long range mediator, the cross-section for the scattering to change the momentum of the blob by $\mathcal{O}\left(1\right)$ is the smaller of $\mu^{-2}$ and the Coulomb scattering momentum transfer cross-section. The latter can be approximated as $1/R_C^2$, where $R_C$ is the classical turn-around point,
\beq
R_C = \frac{\text{W}\left(\frac{g_\chi^2 N_X^2}{\pi v_\chi}\frac{\mu}{M_X}\right)}{\mu} \sim \frac{g_\chi^2 N_X^2}{\pi v_\chi^2 M_X} \,,
\eeq
where $W(x)$ is the Lambert W-function, also known as a product logarithm. Note that the relative velocity is $v_{\chi} \sim 10^{-2}$ in the bullet cluster merger \cite{Robertson:2016xjh}. Additionally, measurements of the Cosmic Microwave Background constrain the energy that can be transferred to the blob in the Early Universe, such that the momentum exchange rate $\frac{1}{M_X v_C}\frac{dE}{dx}$ between the dark matter and the baryons is smaller than the Hubble scale at a redshift factor of $z \approxeq 10^5$~\cite{Dvorkin:2013cea}. Using the calculations of the deposited energy in equations \eqref{eqn:bosededx} and \eqref{Eq:BAODrag}, it is easy to verify that these cosmological bounds are satisfied.\footnote{We note that the expression in \eqref{Eq:BAODrag} applies also to the case of spin dependent interactions in the unpolarized medium of the CMB. This is because the  spin-polarizability of the medium is high, enabling spins to independently respond to the forces created by the blob.} It is important to note that these astrophysical and cosmological bounds apply only if the blobs constituted all of the dark matter. If the blobs were less than $ 10$ percent of the dark matter, the effects of their scattering would not observable in these measurements. Given the complexities of a dark sector with strong self-interactions, it is not unreasonable to expect that only a fraction of the dark matter ends up in blobs within the mass range of interest to us. In our assessment of experimental reach, we will thus consider parts of parameter space where these bounds are violated - it is understood that in this part of parameter space, the expected density of the blobs is $\sim 1/10$ the ambient dark matter density.  

It is also possible to place bounds on these scenarios through the accumulation of dark matter blobs in compact, high density objects such as white dwarfs \cite{Graham:2015apa,Graham:2018efk} and neutron stars \cite{Gould:1989gw}. Over the lifetime of these stars, the accumulated dark matter could trigger explosive processes such as the initiation of runaway nuclear fusion in white dwarfs or the gravitational collapse of the dark matter into a black hole within the star. It is difficult to place model independent bounds on our scenario through these phenomena. The density of the accumulated dark matter in such objects could be much larger than the densities in the blob. This larger density could trigger new processes (through higher dimension operators) within the dark sector (for example, cause a bosenova), causing the destruction of the object well before it accumulates enough matter to affect the star.  Thus, we do not consider potential bounds from these phenomena in this paper.

\section{Fermionic Blob}
\label{sec:fermions}
In this section, we consider the case where the constituents $\chi$ of the blob are fermions. Due to  Pauli exclusion, such a blob has a larger geometric size than the bosonic case. We consider a blob of mass $M_X$ consisting of $N_X = M_X/m_{\chi}$ particles, each of mass $m_{\chi}$. These particles are held together by a strong force $\Lambda_\chi$ and  the blob has a radius $R_X \sim N_X^{1/3}/\Lambda_\chi$. In our analysis, we take $\Lambda_\chi \sim m_{\chi}$ and thus the phenomenology of our blobs should be similar to that of nuclear physics.\footnote{It would be interesting to study the ``atomic'' case, where $\Lambda_\chi \ll m_{\chi}$. We leave this exploration for future work.} In our range 10 keV $\lessapprox \Lambda_\chi \lessapprox$ 10 MeV, the geometric size of the heaviest blobs we consider $M_{X} \sim 10^{33}$ GeV range from $10^{-2}\,$m - $10^2\,$m, scales that are comparable to the dimensions of a lab-scale experiment. It is also straightforward to verify that Fermi degeneracy is sufficient to prevent such blobs from collapsing into black holes. Similar to our analysis of the bosonic blobs, we compute the scattering cross-section between the blob and the standard model in sub-sections \ref{sec:fermionshortrange} and \ref{sec:fermionlongrange}, discuss the bounds on the mediator and  evaluate observational constraints on this kind of blob in sub-section \ref{sec:fermionconstraints}. 

\subsection{Short Range}
\label{sec:fermionshortrange}

Similar to the bosonic case, we assume $\chi$ interacts with the standard model through the Lagrangian: 

\be
\mathcal{L} \supset g_{\chi} \phi \bar{\chi} \chi + g_{N,e} \phi \bar{\Psi}_{N,e} \Psi_{N,e}
\label{eq:fermshortlag}
\ee
Upon integrating out the mediator $\phi$, the effective interaction between $\chi$ and the standard model is again captured by the dimension 6 contact operator 
\beq
\frac{g_{\chi} g_{N,e} \bar{\chi}\chi \bar{\Psi}_{N,e} \Psi_{N,e}}{\mu^2}\, 
\eeq
where $\mu$ is the mass of the mediator. In the fermionic case, since the size of the blob grows with the number of constituents, coherent scatterings are less effective in transferring energy. As discussed in Appendix \ref{sec:BNscattering}, when the size of the blob is larger than the de-Broglie wavelength $\lambda_p$ of the probe, the maximum momentum that can be coherently transferred is $1/\lambda_p$. Due to this limitation, it is important to consider incoherent scattering between the blob and the standard model probe, as well as coherent scattering.

In  incoherent scattering, the momentum transferred can be as large as $\Lambda_\chi$. The energy deposited in a medium of number density $\eta_m$  is 

\be
\frac{dE}{dx}=  \eta_m\, \Lambda_\chi^3\frac{M_X}{m_N} \frac{g_{\chi}^2 g_{N}^2}{\mu^4 v_{\chi}^2}\begin{cases} 1  &\qquad  \Lambda_\chi < m_N v_\chi \\\frac{m^4_N v^4_\chi }{\Lambda^4_\chi} &\qquad  \Lambda_\chi > m_N v_\chi \end{cases}
\label{eq:fermionshortdedx}
\ee
whereas the energy deposited through coherent scattering is 
\be
\frac{dE}{dx} = \eta_m M_{X} \sqrt{\frac{T}{m_N}}\frac{g_{\chi}^2 g_{N}^2\Lambda_\chi^2}{\mu^4 v_{\chi}^2}
\label{eq:fermionshortdedxCoh}
\ee
where $T$ is the temperature of the medium, yielding the de-Broglie wavelength $\sim 1/\sqrt{T m_N}$ of the nuclear probe, and we assume that the geometric cross section is not yet saturated. In both cases, the cross-section is suppressed by phase space factors emerging from the fact that only momenta $\sim \Lambda_\chi$ and $1/\lambda_p$ can be transferred to the nucleus in the incoherent and coherent scattering case respectively. The above formulae are valid in the regime where the scattering cross-section is smaller than the geometric size of the blob---for $\mu \gtrapprox$ TeV, this criterion is satisfied. 

\begin{figure} 
\includegraphics[width=.9\textwidth]{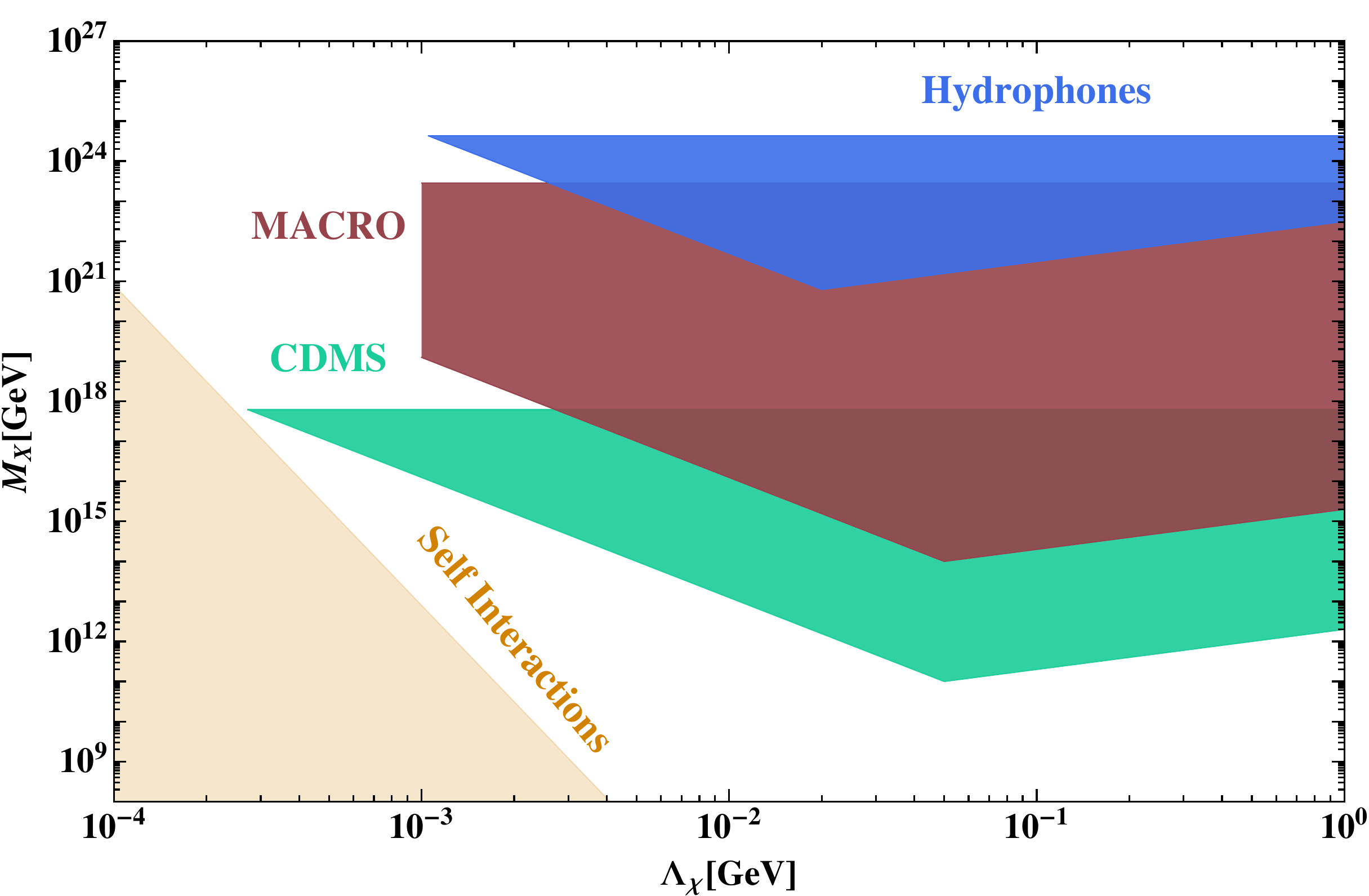}
\caption{$M_X-\Lambda_\chi$ parameter space for a fermionic blob, assuming a short-range mediator of mass $\mu=\,$TeV. Both CDMS and Hydrophones look for total energy deposition while MACRO looks for ionization and scintillation signals.}
\label{fig:FermionShortRange}
\end{figure}

\subsection{Long Range}
\label{sec:fermionlongrange}

Similar to the bosonic case, a fermionic blob sources long range fields that can directly cause accelerations, induce spin precessions and change the values of fundamental constants. We consider the Lagrangian: 
\be
\mathcal{L} \supset g_{\chi}  \phi \bar{\chi}\chi + \underbrace{ g_{N} \phi \bar{\Psi}_{N} \Psi_{N}}_{\CO_1} + \underbrace{\frac{\partial_{\mu}\phi}{f_{N}}  \bar{\Psi}_{N} \gamma^{\mu}\gamma_5 \Psi_{N}}_{\CO_2} + \underbrace{\frac{\phi}{M} F_{\mu\nu}F^{\mu\nu}}_{\CO_3}
\label{eq:fermlonglag}
\ee
where the operator $\CO_1$ leads to accelerations, $\CO_2$ causes spin precessions and $\CO_3$ changes the value of the fine-structure constant.

Much like the bosonic blob, the fermionic blob also sources a classical $\phi$ field. Due to our choice of $\Lambda_\chi \gtrapprox 10$ keV, the bosonic blobs were always physically smaller than the de-Broglie wavelengths of the standard model particles used to probe them. In the fermionic case, the blobs can be significantly larger, $\sim$ $10^{-2}\,$m - $10^2\,$m, for the most massive blobs. For simplicity of analysis, we take the range of the mediator to be longer than the radius of the blob. This long range mediator gives rise to a force that can destabilize the blob. For a repulsive force, the coherent force due to the mediator can decrease the bound state energy such that the blob is no longer stable.\footnote{This mirrors the standard model exactly, where $^{56}$Fe is the most stable nuclei and large nuclei are unstable due to Coulomb repulsion between the protons.} This occurs when the repulsive force $\sim g_{\chi}^2 N_{\chi}^2/R_{\chi}^2$ between two parts of the blob is larger than the attractive surface forces $\sim N_{\chi}^{\frac{2}{3}} \Lambda_{\chi}^2$ holding the two parts together. This leads to the condition  $R_{\chi} \lessapprox \frac{1}{g_{\chi} \Lambda_{\chi}}$. For an attractive force, we demand that the hydrodynamic pressure at the center of the blob is less than $\Lambda_\chi^4$, also yielding  $R_{\chi} \lessapprox \frac{1}{g_{\chi} \Lambda_{\chi}}$. Assuming that the radius of the blob obeys geometric scaling, the dark matter coupling constant is bounded by 
\beq
g_\chi \lessapprox \frac{1}{N_X^{1/3}} \, ,
\eeq
where $N_X$ is the number of constituents in the blob. As with the bosonic case, we demarcate the parts of parameter space that obey these self consistency checks in the sensitivity plots in Figs.~\ref{fig:LIGOFermion}-\ref{fig:Dilaton200km}. Blobs where these conditions are violated may still be found in nature, but they would need additional stabilizing forces. 

This classical field can be used to determine the energy deposition $dE/dx$, spin rotation $\delta \theta$ and strain $h$ induced by the transit of this blob through the standard model, calculated using the impulse approximation techniques of section \ref{sec:bosonlongrange}. A key difference in the phenomenology of the fermionic blob is that, by assumption, the bosonic blob is always smaller than the atomic scale, permitting the entire blob to coherently act on standard model particles during its transit. This is not the case for the fermionic blob---standard model particles can be inside the blob, in which case the force they experience is diminished. The phenomenological change due to this effect is significant - for example, fermionic blobs do not ionize matter as easily, avoiding constraints from the MACRO experiment.

\subsection{Constraints}
\label{sec:fermionconstraints}

The bounds on the mediator depend solely on the standard model - thus they are the same for the bosonic and fermionic cases. The key difference between the bosonic and fermionic case is that the physical size of the fermionic blob grows with its mass. As in the bosonic case, we take the short range mediator to have a mass $\gtrapprox$ TeV and restrict ourselves to long range mediators in the mass range  $\mu\text{m}^{-1} - \left( \text{6000 km}\right)^{-1}$. For the long range case, to simplify our analysis, for a given blob we only consider mediators whose range is longer than the size of the blob.

The physical size of the blob dilutes constraints that rely on localized energy deposition such as MACRO since standard model particles are subjected to a smaller force from the blob.  On the other hand, bounds from the bullet cluster can get stronger since the geometric size of the blob is now larger. The self scattering cross-section per unit mass ($\sigma/M$) of the fermionic blob is $\sim 1/\left(M_{X}\Lambda_{\chi}^8\right)^{\frac{1}{3}}$. Unless the blob is a sub-dominant component of dark matter, the bound from the bullet cluster requires
\beq
M_{X} \gtrapprox 10^{40} \text{ GeV} \left( \frac{\text{10 keV}}{\Lambda_{\chi}} \right)^{8} \left(\frac{ 1 \text{ cm}^2/\text{gm}}{\sigma/M}\right)^{3}\, .
\eeq This bound is independent of any additional long range interaction between the blobs.

\section{Detection Methods}
\label{sec:detection}
The transit of the blob is a rare event due to its low number density. However, the accumulation of dark matter in the blob allows for these rare transits to cause observable transients in terrestrial detectors. A search for these signals requires methods to distinguish it from backgrounds. There are two potential handles that could be exploited to achieve this goal. First, the dark matter moves with a speed $\sim 220$ km/s, significantly faster than any terrestrial source of noise, but significantly slower than the speed of light, placing it in a unique range of speed between terrestrial and cosmic ray induced events. If the signal from the dark matter is large enough to be observed at multiple locations in a detector that also has sufficient temporal resolution, it should be possible to distinguish this signal from other background transients. These events should also lie along a straight line, enabling further background rejection.   Second, the dark matter has the ability to pierce through shields and interact in its own unique way with standard model sensors. Thus, in a setup that is monitored with a variety of precision sensors, the collective information from all sensors could potentially be used to reject standard model backgrounds. This latter option is technically challenging, but it is similar in spirit to WIMP detection experiments that use data from multiple channels to veto standard model events. Similar protocols could also be employed in experiments such as LIGO which monitor a variety of potential noise sources. 

In the following, we describe the reach of current and proposed detectors to the transient signals caused by dark matter blobs. These estimates are made using the statistical sensitivity of the detectors, assuming that systematics can be combated with the above handles. 

\subsection{Ionization}
\label{subsec:ionization}

Energy depositions  $\sim$ eV/\AA $\,$ that cause ionization are constrained by the MACRO experiment~\cite{Ahlen:1992pe, Ambrosio:2002mb, Ambrosio:2002qq}. At weaker coupling, multiple scattering events are still possible in a detector.  These scattering events can occur through the collision of the blob with nuclei, depositing energy in detection channels often searched for in conventional WIMP detection experiments~\cite{Akerib:2016vxi,Aprile:2017iyp,Aprile:2018dbl}. This is similar in spirit to recent CDMS searches for lightly ionizing particles~\cite{Agnese:2014vxh}.

\begin{figure} 
\includegraphics[width=.9\textwidth]{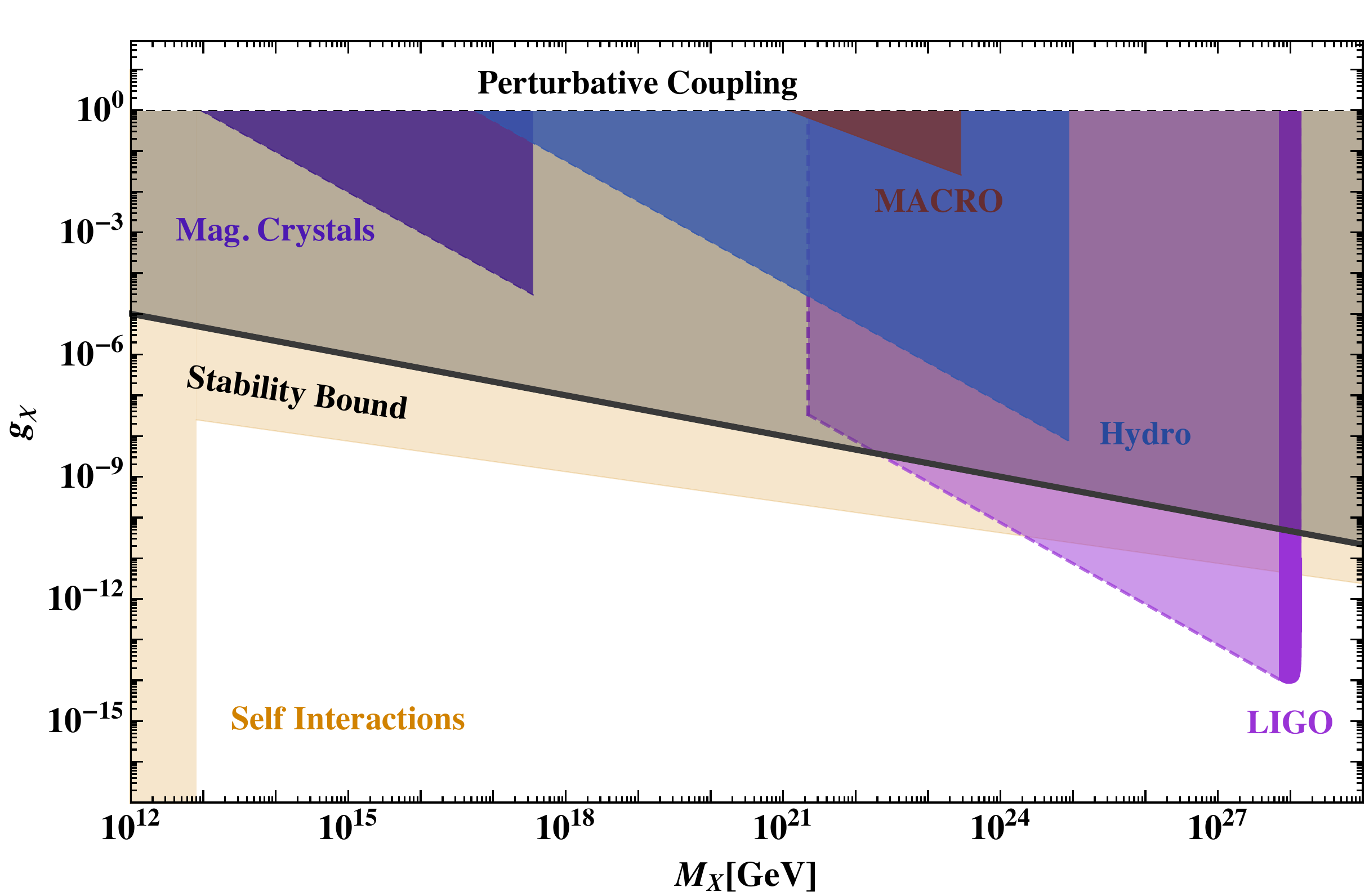}
\caption{Reach on coupling $g_\chi$ for a scalar mediator of range 20 km coupling to a fermionic blob with $\Lambda_\chi=1\,$MeV, as a function of blob mass $M_X$. The different shading for the reach of both interferometer experiments is explained in Sec.~\ref{sec:discussion}.}
\label{fig:LIGOFermion}
\end{figure}

\begin{figure} 
\includegraphics[width=.9\textwidth]{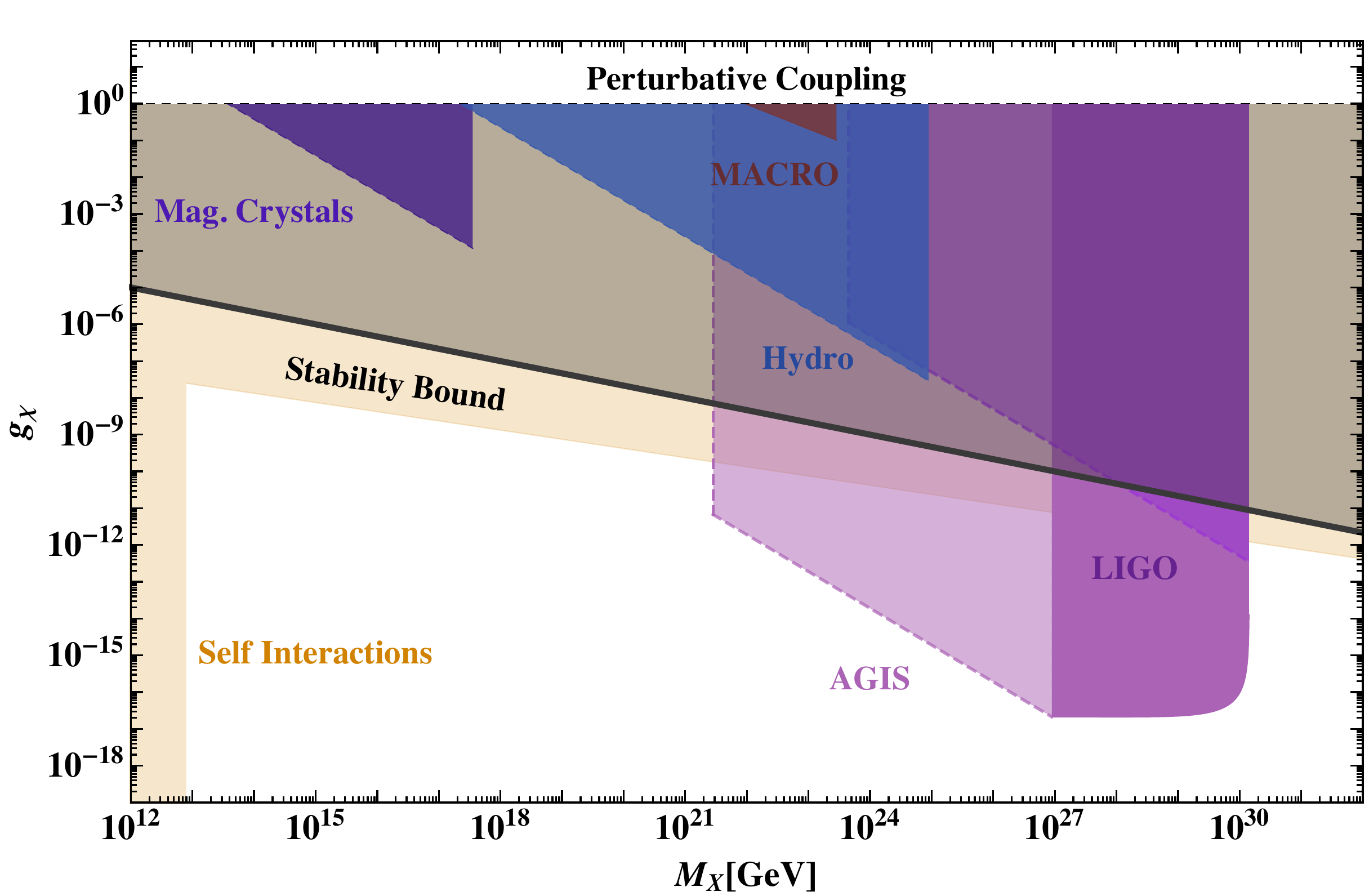}
\caption{Reach on coupling $g_\chi$ for a scalar mediator with a 200 km range coupling to a fermionic blob with $\Lambda_\chi=1\,$MeV, as a function of blob mass $M_X$. The different shading for the reach of both interferometer experiments is explained in Sec.~\ref{sec:discussion}.}
\label{fig:AGISFermion}
\end{figure}

\subsection{Acoustics}
\label{subsec:acoustics}

Collisions between nuclei and soft blobs (with $\Lambda_\chi \lessapprox$ MeV) can lead to significant energy depositions, even though the deposited energy is too small to cause ionization. The heat deposited in such a collision could potentially be detected using two different techniques. First, the localized heat deposition can produce phonons/sound waves that could be detected using sensitive acoustic detectors.  Such a signal would be visible in acoustic detectors such as CDMS where the total energy deposited in the detector is large enough to enable calorimetry. For example, energy depositions $\sim$\,keV/cm are measurable in CDMS's calorimeters. Since these events do not cause ionizations, traditional techniques cannot be used to distinguish these events from noise. However, the transit of the blob will lead to a line of hot cells, which should enable signal recognition. These sound waves could also potentially be searched for in networks of hydrophone sensors, particularly for ultra-heavy blobs ($M_{X} \gtrapprox 10^{20}$ GeV) that require large volume detectors. A quick estimate of the reach of hydrophone networks can be made using the formalism of \cite{Learned:1978iv} which shows that energy depositions dE/dx $\sim 10$ keV/\AA \,  spread within an area $\lessapprox \left(1 \text{ mm}\right)^2$ can be observed with hydrophones with a sensitivity $\sim\,10^{-4} \text{ dynes/}\left(\text{cm}^2 \text{Hz}\right)$ at frequencies $\sim$ 100 kHz out to distances $\sim 100$\,m. We leave a detailed analysis of this detection method for future work \cite{SRandNaoko}. The estimated reach is shown in Fig.~\ref{fig:FermionShortRange}.

The second possibility would be to make use of in-situ amplification where the localized energy deposition leads to an amplified material response. Single-molecular magnets~\cite{Bunting:2017net}, where heat deposition ($\sim$ 10 meV/\AA) triggers an amplified magnetic avalanche, have recently been identified as possessing properties favorable for low threshold dark matter detection. If successfully developed, these detectors can also search for these soft collisions.

\subsection{Acceleration}
\label{subsec:acceleration}

The blob can cause an acceleration on a test body through the long range force exerted by it or through direct collisions with the body as it transits through it. In the case of a long range force, the force is effective when the dark matter is within the range $\mu^{-1}$ of the force. The test body will respond freely to this force for a time that is the shorter of the transit time $\sim 1/\left( \mu v_{\chi}\right)$ and the period $t_r$ of the restoring forces supporting the test body ({\it i.e.} the time period $\sim$ 0.025 s of LIGO's mirrors, the free fall time $\sim 1$ s in an atom interferometer). The displacement during this time should be compared to the position sensitivity of the sensor integrated over the transit time to obtain the reach of the experiment. For concreteness, we estimate the reach of LIGO~\cite{0264-9381-32-7-074001} with a position sensitivity of~$ \CO(10^{-17} \text{m/}\sqrt{\text{Hz}})$ (around 100 Hz) in Figs.~\ref{fig:LIGOFermion} and ~\ref{fig:AGISFermion} and the reach of AGIS~\cite{Dimopoulos:2008sv} with a position sensitivity of $\CO(10^{-18}\text{m}/\sqrt{\text{Hz}}
)$ in Fig.~\ref{fig:AGISFermion} (around 1 Hz).   

For short range interactions, the energy deposited by the blob during its transit is largely expended into phonons. This energy is not detected by accelerometers, which are sensitive to the center of mass displacement of the test body. For a test body of mass $M$ and length $L$, the energy deposited in the transit is $\sim(dE/dx)L$ leading to a displacement $\sim \sqrt{\left(dE/dx\right)L/M} (L/v_{\chi})$. In order to be detectable at LIGO whose mirrors have a mass $M \sim$ 40 kg and $L \sim 10$ cm, the energy deposition needs to be $dE/dx \gtrsim \text{MeV/cm}$. LIGO is thus not as sensitive to these collisions as dedicated calorimeter experiments such as CDMS.

\subsection{Spin Precession}
\label{subsec:spinprecession}

\begin{figure}
\includegraphics[width=.9\textwidth]{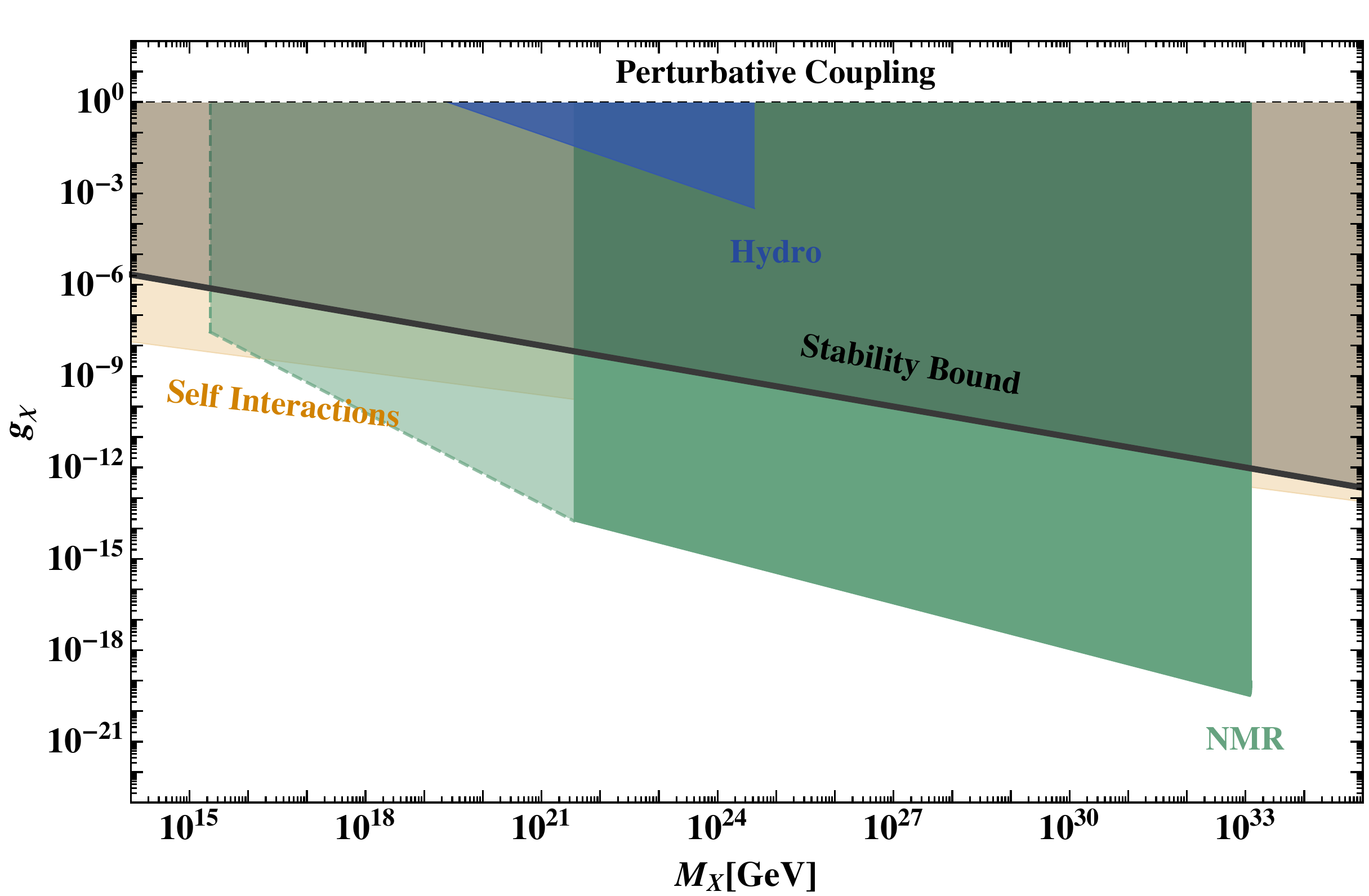}
\caption{Reach on coupling $g_\chi$ for a fermionic blob with $\Lambda_\chi=\,$MeV and a $6000\,$km-range pseudoscalar mediator, as a function of blob mass $M_X$. The different shading for the reach of NMR-type experiments is explained in Sec.~\ref{sec:discussion}.}
\label{fig:NMRFermions}
\end{figure}

The classical field from the blob can induce a torque on nucleon and electron spins causing them to precess. This precession changes the magnetization of a sample and can be measured using precision magnetometers such as a SQUID. The spins will freely precess for a time that is the shorter of the transit time $1/\left( \mu v_{\chi}\right)$,  the spin relaxation time $T_2$ and the Larmor period of the sample. Since electron spins at high density have short $T_2$ relaxation times, it is advantageous to search for these effects in nucleons. Of particular interest are spin precession experiments using liquid Xenon (such as CASPEr-Wind~\cite{Budker:2013hfa}) where $T_2 \approx 1000$ s have been demonstrated. Moreover, CASPEr-ZULF~\cite{JacksonKimball:2017elr} has also demonstrated the capability to operate NMR experiments in Liquid Xe at zero/ultra-low magnetic fields giving rise to Larmor periods as long as one second. In this paper, we limit the range of the mediator to be $\lessapprox$ 6000 km giving rise to transit times $\sim$ 10 s. Since this is within the range of estimated CASPEr-ZULF capabilities, we will assume that the spin precession is limited by the transit time of the blob. The sensitivity of the experiment to the blob is then estimated by demanding that the change in the sample magnetization is larger than the noise in the SQUID ($\sim 0.1 \text{fT}/\sqrt{\text{Hz}}$) integrated for the transit time $\sim 1/\left(\mu v_{\chi}\right)$. The estimated reach in an approximentaly $\left(10 \text{ cm}\right)^3$ Liquid Xe sample is shown in Fig.~\ref{fig:NMRFermions}. 

\subsection{Strain}
\label{subsec:strain}

\begin{figure}
\includegraphics[width=\textwidth]{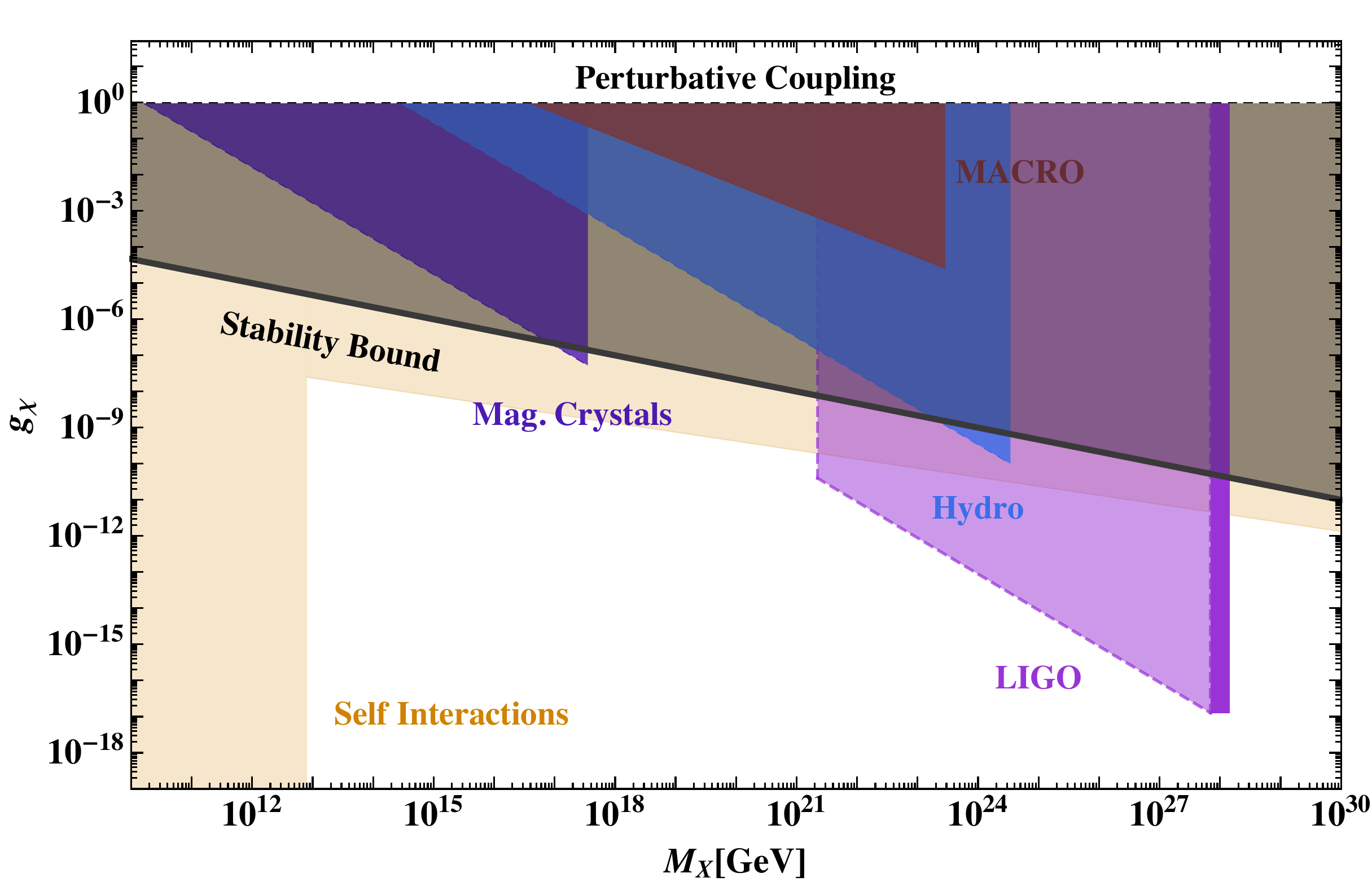}
\caption{Reach on coupling $g_\chi$ for a dilaton mediator with 20 km range for a fermionic blob with $\Lambda_\chi=1\,$MeV, as a function of blob mass $M_X$. The different shading for the reach of both interferometer experiments is explained in Sec.~\ref{sec:discussion}.}
\label{fig:Dilaton20km}
\end{figure}

\begin{figure}
\includegraphics[width=\textwidth]{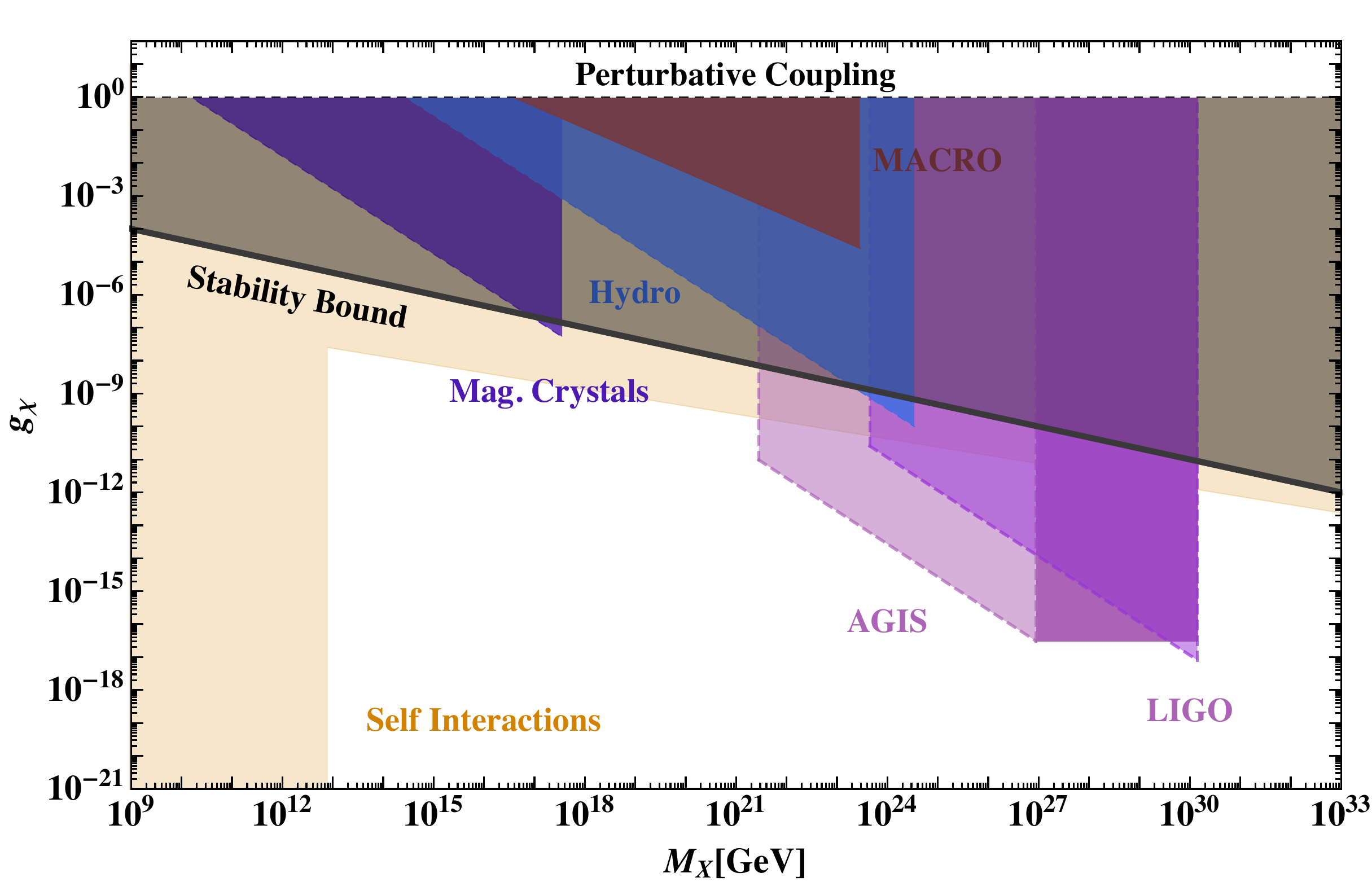}
\caption{Reach on coupling $g_\chi$ for a dilaton mediator with range of 200 km for a fermionic blob with $\Lambda_\chi=1\,$MeV, as a function of blob mass $M_X$. The different shading for the reach of both interferometer experiments is explained in Sec.~\ref{sec:discussion}.}
\label{fig:Dilaton200km}
\end{figure}

The blob can source dilatonic/moduli fields that directly change the values of fundamental constants such as the mass and charge of the electron. For concreteness, we consider a charge modulus. The transit of such a blob can directly exert forces on standard model particles, leading to accelerations that can be measured using experiments such as LIGO, as discussed in \cite{Hall:2016usm}.  The estimated reach for these forces is shown in Figs.~\ref{fig:Dilaton20km} and~\ref{fig:Dilaton200km}. There is however a more direct effect. Changes to fundamental constants leads to shifts in atomic energy levels/Bohr radii causing shifts in the lengths and transition frequencies of physical systems. These can be measured in gravitational wave bar detectors~\cite{Astone:1993ur,Astone:1997gi}, LIGO~\cite{0264-9381-32-7-074001} and atomic clock systems~\cite{ Dimopoulos:2008sv}.

For simplicity, we make estimates only for systems where the range $\mu^{-1}$ of the modulus is longer than the length of the sensing apparatus. For a bar detector, the strain Eq.~\eqref{Eq:Strain} leads to a direct change in the length of the bar. In LIGO, the change to atomic transitions will change the frequency of the output laser. But, this effect is common to both arms of the interferometer and is thus canceled in the differential measurement. However, the physical length of LIGO's arms will also change due to the modulus. The gradient in the sourced modulus field causes these arm lengths to change differently, leading to a measurable signal. This reach is plotted in Figs.~\ref{fig:Dilaton20km} and \ref{fig:Dilaton200km}.  In atomic clock systems, there are two effects. First, two local clocks whose transitions depend upon different powers of $\alpha_\text{EM}$ can be compared. The blob will shift the energies of these transitions differently causing a signal.  Second, the modulus field sourced by the blob will change atomic transitions differently over a baseline. This leads to a signal in single-baseline atomic gravitational wave detectors such as AGIS. In all of these cases, the reach is estimated by comparing the signal with the noise in the detector at a period equal to the transit time of the blob. This reach is plotted in Fig.~\ref{fig:Dilaton200km}.

\section{Results and Discussion}
\label{sec:discussion}

\subsection{Bosonic Blob Reach}
We begin by exploring the parameter space of a bosonic blob, first considering the case of a short-range mediator between $\chi$ and the standard model, resulting in the contact operator given in Eq.~\eqref{eq:BosShortRangeOperator}. In Fig.~\ref{fig:BosonShortRange} we show the region of $g_\chi$, defined through Eq.~\eqref{eq:bosshort}, accessible as a function of mediator mass $\mu$, for a blob of mass $10^{16}\,$GeV and a Bohr radius $\Lambda_\chi=100\,$keV. We consider the case $g_N=1$, whence the cross section, Eq.~\eqref{eq:shortbosxsec}, saturates the geometric cross section bound, {\it i.e.} $\sigma \sim \Lambda_\chi^{-2}$. For a Bohr radius of $\Lambda_\chi = 100\,$keV, the blob cannot impart enough momentum to a nucleus to cause ionization; for higher values of $\Lambda_\chi$, the total energy deposition per unit length is below the detector sensitivity of MACRO. Therefore, only the CDMS calorimeters are sensitive to  bosonic blobs interacting only via short range interactions, at least in this blob mass range.  

Moving next to long range mediators, we consider three different possible interactions between the blobs and the standard model, as described by the operators $\CO_1$ (scalar), $\CO_2$ (pseudoscalar) and $\CO_3$ (dilaton) in Eq.~\eqref{eq:boslonglag}, each of which can be probed using the novel experimental searches described above. Additionally, energy deposits arise in all three scenarios, which require qualitatively different search strategies depending on whether they are ionizing or not.  In detailing the reach of various experiments, we maximize the parameters $g_N$, $f_N$, and $M$ that appear in this Lagrangian to a value consistent with the existing experimental constraints, described in Sec.~\ref{subsubsec:bos5thforce}: $g_N$  takes values $7\times 10^{-21}$  and $2\times 10^{-21}$ for mediator ranges of $20\,$km and $200\,$km, while  $f_N=5\times 10^9\,$GeV and $M=10^{14}\,$GeV. 

For the case of a scalar mediator, interacting via $\CO_1$, we consider parameter space in the $g_\chi-M_X$ plane, shown in 
Fig.~\ref{fig:AGISBoson} for a Bohr radius $\Lambda_\chi=10\,$keV, and a mediator range of $200\,$km. The parameter regions labeled Mag. Crystals and MACRO show the sensitivity of those experiments to the energy deposited in them during a blob transit, as calculated by Eq.~\eqref{eqn:bosedynfric} with the force $F(r)$ that of a scalar-scalar interaction, convolved over the blob. For the magnetic crystals we estimate sensitivity to a $dE/dx=10^{-3}\,$eV per angstrom, and require one such event per year; for MACRO we require an energy  of $10\,$ eV to be imparted to an individual nucleus per unit length of the detector, such that ionization occurs, and require one transit per 10 years (the runtime of the experiment). The purple regions labeled LIGO and AGIS show the sensitivity of these experiments to the acceleration of the test particles caused by the transit (Sec.~\ref{subsec:acceleration}). Here, the gradient in the field over the experiments causes the test masses to move different distances during the transit time of the blob. Note that for LIGO, the gradient is $\sim$(4\,km/200\,km), the mass of test mass is 40 kg and the period of the test mass is 1/40 seconds; for AGIS, the gradient is $\sim$(1\,km/200\,km) and the test mass is $\sim$\,90 GeV. To estimate the reach, we require at least one event per year and that the relative movement of the test masses be that of the position sensitivity of the experiment at a frequency of approximately the inverse transit time. 

For light blob masses, we introduce an artificial cut-off to simply the reach analysis. The lighter shaded region enclosed by a dashed line denotes that the reach includes an artificial cut-off: we demand that the blob does not pass through the detector, even though for such masses, there can be multiple blobs that pass through the detectors every year.  Blobs passing closer to the detector, particularly inside the detector itself, would result in a more complicated signal than the one we estimate here, dependent on the detector geometry; we expect a detailed study of such events would lead to improved sensitivity in this parameter region, but this is beyond the scope of the current exploratory work. The lower bound on the blob masses is due to the requirement that there only be one blob passing close to the detector during the experimental integration time, in order to have an unambiguous signal. Similarly, multiple blobs interacting with the detector at the same time would lead to a more complicated signal with a potentially better handle on background discrimination, but a study of such events is beyond the scope of this work. The artificial cutoffs for a mediator with a 200 km range are 5 km and an integration time of 100 s for AGIS and slightly less than 200 km and a 10 s integration time for LIGO. The stability bound, Eq.~\eqref{eq:bosstab}, and the bounds on self-interactions described in Sec.~\ref{subsubsec:boscosmo} are also shown.

For the case of a pseudoscalar mediator, interacting via $\CO_2$, we show in Fig.~\ref{fig:NMRBosons} the estimated sensitivity in $g_\chi-M_X$ parameter space for a blob with Bohr radius $\Lambda_\chi=10\,$keV and a mediator with range $R \sim 6000\,$km, the radius of the Earth. Again the regions labeled Mag. Crystals and MACRO show the sensitivity of these experiments to energy deposits (we require the same $dE/dx$ per blob transit as in the scalar case). In this case, the standard model interaction with the mediator is via a dipole force, which enters into Eq.~\eqref{eqn:bosedynfric}. This force can be calculated using the potential of the standard model spin $\mathbf{m}$ in an effective magnetic field sourced by the blob 
\beq
\mathbf{B}_{\text{eff}}\sim g_N g_\chi N_\chi\left(-\frac{\mu}{r}+\frac{1}{r^2}\right)  e^{-\mu r} \,\,\hat{ \mathbf{r}} 
\eeq 
as $\mathbf{F} = -\nabla(\mathbf{m}\cdot \mathbf{B}_{\text{eff}})$. The green region labeled NMR estimates the sensitivity of the CASPEr-ZULF experiment to the spin precession caused by the blob transit, as described in Sec.~\ref{subsec:spinprecession}. Using Eq.~\eqref{eq:deltheta} to obtain the angle shift of the spins during a transit, we compare the resulting shift in magnetization for a spin density of $10^{22}\,\text{cm}^{-3}$, and require that this shift is observable within a spin relaxation/Larmor time of $10\,$s, assuming a sensitivity of 0.1\,fT/$\sqrt{\text{HZ}}$. As above, the lighter shaded region with dashed borders indicates a (conservative) artificial cut-off in our estimate: we require that the blob does not transit within a region 10m around the detector so we can ignore detailed geometry and that there be only one blob interacting with the detector during a period of twice the spin relaxation time. The stability and self-interaction bounds are calculated as described above for the scalar mediator case.

Finally, for the case of a dilaton mediator, interacting via $\CO_3$, we show in Fig.~\ref{fig:DilatonBosons} the estimated sensitivity in $g_\chi-M_X$ parameter space for a blob with Bohr radius $\Lambda_\chi=10\,$keV and a mediator with range $200\,$km. In this case, the force that enters into the calculation of the $dE/dx$ sensitivities of magnetic crystals and MACRO is that induced from the change in the electromagnetic contribution to the nuclear mass, Eq.~\eqref{eq:emmassforce}. The purple curves labeled LIGO and AGIS show the parts of parameter space accessible to these experiments via the mechanism described in Sec.~\ref{subsec:strain}. The sensitivities were estimated by equating the strain due to the shift in bond length induced by the variation of $\alpha_\text{EM}$, Eq.~\eqref{eq:delh}, to the experimental position sensitivity and requiring at least one such event per year. The lighter dashed regions again denote regions with our imposed veto on blobs passing with a certain distance of the detector and during a certain integration time; these cut-offs are the same as for the scalar case described above. The stability and self-interaction bounds are calculated as described above for the scalar mediator case.

\subsection{Fermionic Blob Reach}

We begin our analysis of the fermionic blob by studying the case of the short-range mediator, described by the Lagrangian Eq.~\eqref{eq:fermshortlag}. In Fig.~\ref{fig:FermionShortRange} we fix the mass of the mediator $\mu=\,$TeV, fix $g_\chi=g_N=1$, and study the $M_X-\Lambda_\chi$ parameter space. We estimate the sensitivity to energy deposits in MACRO, CDMS, and a hydrophone experiment, as described in Sec.~\ref{subsec:ionization} and Sec.~\ref{subsec:acoustics}. The region of parameter space for CDMS and hydrophones is bounded from above by requiring one event per year in the experimental volume--we assume a hydrophone tank of (500\,m)$^3$; for MACRO we require one event in the 10 year run time. We calculate the energy deposited, $dE/dx$, using formula Eq.~\eqref{eq:fermionshortdedx}, valid  for the range shown, $10^{-4}\,\text{GeV}\lesssim\Lambda_x\lesssim1\,$GeV. Requiring this energy deposit to be above keV/cm and 10 keV/angstrom for CDMS and hydrophones, respectively, results in the negative slope that bounds their sensitivity region from the left; for MACRO, as in the bosonic case, we require a $dE/dx \sim$ MeV/cm, as well as an energy deposition into individual nuclei that is sufficient to ionize, resulting in a sharp vertical cutoff at $\Lambda_\chi\sim1\,$ MeV. The `V' shape of the boundary comes from the fact that the maximum transfer is the minimum of $\Lambda_\chi$ and the momentum of the standard model probe, $m_N v_\chi$, as detailed in Eq.~\eqref{eq:fermionshortdedx}. Note that the energy deposition is mostly due to incoherent scattering. The self-interaction constraints, discussed in Sec.~\ref{sec:fermionconstraints}, are also shown. 

Turning to the case of a long-range mediator, with Lagrangian given in Eq.~\eqref{eq:fermlonglag}, our calculation of the viable parameter regions mirrors that of the bosonic case --- as mentioned, the only technical difference comes from the size of the fermionic blob. This affects the various calculations of energy deposit in each experiment, described above for a bosonic blob in the case of a scalar, pseudoscalar, and dilaton mediator, entering the calculation of $F(r)$ in Eq.~\eqref{eqn:bosedynfric}, which is a superposition of the individual classical fields sourced by each $\chi$ within the blob. For example, for a scalar mediator, the total energy deposition, $\Delta E_\text{tot}$ is related to maximum energy transferred to a single standard model probe, $ \Delta E_\text{max}$ by
\beq
\Delta E_\text{tot} \approx \left(\frac{R_X}{R_0}\right)^2 \Delta E_\text{max}
\eeq
where $R_X$ is the radius of the blob and $R_0$ is the radius of the standard model atom, approximately an Angstrom. This approximate relation results from the observation that all the standard model probes that pass within the radius of the blob receive approximately the same momentum kick and so the total energy deposition per unit length is enhanced by the total number of probes that pass through the blob.  Modulo this change, which can be implemented numerically, all calculations proceed as in the bosonic blob case. We note that this modification of  $dE/dx$ weakens the bounds from MACRO, which requires ionization. We also include in the fermionic case the projected sensitivity from energy deposits in a hydrophone-rigged ($500$\,m)$^3$ tank of water, requiring one event per year with  $dE/dx=$10 keV/angstrom. The plots we show are the parameter regions in the $g_\chi-M_X$ plane for constituents of mass $\Lambda_\chi=1\,$MeV. Note that we use the same shading convention as in the bosonic case. The artificial cutoffs are as follows:  14km and 10s for the LIGO sensitivity reach in Figs.~\ref{fig:LIGOFermion} and Fig.~\ref{fig:Dilaton20km}, 5 km and an integration time of 100 s for AGIS and slightly less than 200 km and a 10 s integration time for LIGO in Figs.~\ref{fig:AGISFermion} and Fig.~\ref{fig:Dilaton200km} and 10 m with a 20 s integration time for Fig.~\ref{fig:NMRFermions}.

For all types of mediator, and for both the bosonic and fermionic blob, we find interesting and experimentally accessible parameter space.

\section{Conclusions}

Self interactions can cause the dark matter to accumulate into large composite states necessitating new experimental search strategies. In this paper, we have argued that the challenge of detecting the small number density of these composite states can potentially be overcome in current (XENON, LUX, CDMS and LIGO) and planned (CASPEr, AGIS) detectors by leveraging the fact that the speed of the dark matter lies in a unique window between relativistic and terrestrial sources of noise. The enhanced interaction of composite states with the standard model could lead to the excitation of multiple detector modules. In concert with precision timing, these excitations can be used to reject noise and identify the dark matter signal. These signatures can be identified either through changes to the data analysis protocol or through relatively straightforward changes to the way the experiments are run. In this paper, we identified regions of parameter space that are consistent with current observational bounds and can be experimentally probed. This is by no means an exhaustive study - our goal was to simply establish experimentally interesting targets. While there are many theories of composite dark matter, we have categorized their experimental signatures, enabling a systematic probe of this parameter space.

There are several aspects of the phenomenology of blobs that deserve further study. For example, we have adopted an agnostic approach towards the production of these blobs. Exploration of the production of  blobs in the early universe has focused on the case when the blobs primarily interact with each other through contact interactions. In these cases, the low velocity of the blob inhibits its subsequent growth. It would be interesting to explore the evolution of these blobs in the galaxy wherein they acquire significant virial velocities potentially enabling an additional period of blob growth. Moreover, long range forces in the dark sector could also lead to further growth in the dark sector. These additional interactions could potentially resolve the tensions of the cold dark matter paradigm, particularly in the regime of small scale structure.  

In this paper, we focused on terrestrial direct detection experiments.  It would be interesting to explore the qualitatively different characteristics of dark blobs in indirect detection experiments (see Ref.~\cite{Gresham:2018anj} for some recent results in this direction)---for example, the collision of dark blobs with each other would release a large number of high energy particles in rare, localized bursts. Such collisions are likely to exhibit different spatial and temporal statistical features than the conventional expectations from annihilating and decaying dark matter. The entropy released in these collisions could also lead to qualitatively new kinds of cosmic rays - for example, such collisions might produce complex anti-particle nuclei such as anti-deuterium that are rarely produced by standard model or conventional dark matter scenarios. These indirect detection signatures are particularly relevant in searching for blobs with mass greater than $10^{33}$ GeV, wherein the flux of the blob is too small to transit through the Earth. In addition to cosmic ray signatures, these collisions could also cause new astrophysical phenomena (see Refs.~\cite{Hardy:2015boa,Gresham:2018anj}) such as triggering the explosion of sub-Chandrasekhar mass white dwarfs \cite{Graham:2015apa,Graham:2018efk}.

\begin{acknowledgments}
We wish to thank Peter Cox, David E. Kaplan, Simon Knapen, Tongyan Lin, Tim Lou, Dan McKinsey, Matt Pyle, Katelin Schutz, and Kathryn Zurek for useful discussion over the course of the this work. Part of this work was performed at the Aspen Center for Physics, which is supported by National Science Foundation grant PHY-1607761; DMG thanks the Aspen Center for Physics and Kavli IPMU for the hospitality shown while this work was being completed. DMG is funded under NSF Grant 32539-13067-44--PHHXM and DOE Grant 041386-002.  TM acknowledges support from KITP while this work was being completed, which is supported in part by the National Science Foundation under Grant No. NSF PHY-1748958. TM is supported by WPI, MEXT, Japan.  SR was supported in part by the NSF under grants PHY-1638509 and PHY-1507160, the Alfred P. Sloan Foundation grant FG-2016-6193 and the Simons Foundation Award 378243.   This work was supported in part by Heising-Simons Foundation grant  2015-038.
\end{acknowledgments}

\appendix

\section{Blob-Nucleon Scattering}
\label{sec:BNscattering}

Scattering between a point particle and a composite object is a well known quantum mechanical problem, assuming that the point particle may be treated as free. In this case, the initial wavefunction is  a plane wave and one may calculate the cross section using standard methods. 
In this Appendix we present a quantum mechanical treatment that takes into account the localized extent of the point particle's wavefunction---in our case the standard model nucleus---paying particular attention to how this alters the coherent enhancement of the differential cross section at low momentum transfer.

We begin by recalling the standard scattering story in quantum mechanics, following e.g. Sakurai~\cite{Sakurai:1167961}. Define $| \phi \rangle$ as a plane wave eigenstate of free particle Hamiltonian $H_0$ obeying
\beq
H_0 |\phi \rangle = E |\phi \rangle \,,
\eeq
and $| \psi \rangle$ as an eigenstate of the full Hamiltonian, obeying
\beq
\left(H_0 + V\right) |\psi \rangle = E |\psi \rangle \,,
\eeq
with $V$ being a time-independent potential. The Lippmann-Schwinger equation provides a solution $|\psi^\pm\rangle$ to the above such that $|\psi\rangle\to|\phi\rangle$ as $V\to0$,
\beq
|\psi^\pm \rangle =|\phi \rangle+ \frac{1}{E - H_0 \pm i \epsilon}V |\psi^\pm \rangle \, .
\label{eq:LSFull}
\eeq
If $H_0$ is the free Hamiltonian,  this can be written in the wavefunction representation as
\beq
\psi(\mathbf{x}) \sim \left(\frac{1}{2\pi}\right)^{3/2}e^{i \mathbf{x}\cdot \mathbf{k}} - \frac{m}{2\pi r}\int d^3 x' \, e^{- i  \mathbf{x}'\cdot \mathbf{k}'}V(\mathbf{x}') \psi(\mathbf{x}) \qquad \qquad \mathbf{k}' = |\mathbf{k}| \frac{\mathbf{x}}{r}
\label{eq:LippmannSchwinger}
\eeq
where $r = |\mathbf{x}|$. The scattering cross section is extracted from Eq.~\eqref{eq:LippmannSchwinger} by casting it in the form 
\beq
\psi(\mathbf{x}) \sim \left(\frac{1}{2\pi}\right)^{3/2}\left(e^{i \mathbf{x}\cdot \mathbf{k}} + f(\mathbf{k}', \mathbf{k}) \,\frac{ e^{i k r}}{r}\right) \,,
\eeq
where $f(\mathbf{k}', \mathbf{k})$ is the scattering amplitude, viz.
\beq
\frac{d \sigma}{d \Omega} = \left|f(\mathbf{k}', \mathbf{k})\right|^2
\eeq
In the Born approximation, where the potential is sufficiently weak, the scattering amplitude is
\beq
f^{(1)}(\mathbf{k}, \mathbf{k}') = -\frac{m}{2\pi} \int d^3x' \, e^{i  \mathbf{x}'\cdot \left(\mathbf{k}-\mathbf{k}'\right)}V(\mathbf{x}')
\eeq
and can be explicitly calculated for a multitude of scattering potentials. 

To see the origin  of coherent enhancement at low momentum transfer, take as an example, a potential arising from multiple scattering centers,
\beq
V(\mathbf{x}') = \sum_{i = 1}^{N_S} v(\mathbf{x}' - \mathbf{x}_i)
\label{eqn:MultPot}
\eeq
where $v(\mathbf{x})$ is the potential due to a single scattering center and $\mathbf{x}_i$ are the locations of the $N_S$ scattering centers. Assume that the total scattering center is localized within a radius $R$. For this potential, the scattering amplitude is 
\beq
f^{(1)}_{\text{free}}(\mathbf{k}, \mathbf{k}') &=& -\frac{m}{2\pi} \int d^3x' \, e^{i  \mathbf{x}'\cdot \left(\mathbf{k}-\mathbf{k}'\right)}\sum_{i = 1}^N v(\mathbf{x}' - \mathbf{x}_i) \nonumber \\
&=& -\frac{m}{2\pi} \sum_{i = 1}^{N_S}  \int d^3x' \, e^{i  \mathbf{x}'\cdot \mathbf{q}}v(\mathbf{x}' - \mathbf{x}_i)
\eeq
where $\mathbf{q}=\mathbf{k}-\mathbf{k}'$ is the momentum transfer. When the momentum transfer obeys $q R \ll 1$, the exponential is essentially unity and
\beq
f^{(1)}_{\text{free}}(\mathbf{k}, \mathbf{k}') &=& N_S \,f^{(1)}_1
\eeq
where $f^{(1)}_1$ is the scattering amplitude for a single scattering center. Therefore, the differential cross section, in this limit, is
\beq
\frac{d\sigma}{d\Omega} &\sim& N_S^2\, \frac{d\sigma_1}{d\Omega}
\eeq
where $d\sigma_1/ d\Omega$ is the differential cross section for a single scattering center. When $q R \gtrsim 1$, the phases for the different scattering centers cancel and so the cross section is incoherent, {\textit i.e.}
\beq
\frac{d\sigma}{d\Omega} &\sim& N_S\, \frac{d\sigma_1}{d\Omega} \, .
\eeq

The above treatment is valid even when the point particle is not free, so long as the size of the particle's wavepacket is larger than any other relevant length scale in the problem. This is usually true in most scattering situations. However, when a dark blob is much larger than an Angstrom one has to take into account the localization of the standard model particle's wavefunction. 

Simple reasoning provides a heuristic picture as follows. Let us take  $H_0$  to now include the potential of an infinitely deep three dimensional square well, so as to idealize the localization of the wavefunction of the standard model particle. In this case, 
\beq
H_0 = -\frac{1}{2m}\nabla^2 + V_W(\mathbf{x}) \qquad \qquad V_W(\mathbf{x}) = \begin{cases} 0 &\qquad |x_i|< \frac{L}{2} \nonumber \\ \infty &\qquad |x_i|> \frac{L}{2} \end{cases}
\eeq
The Lippmann-Schwinger equation, eq.~\eqref{eq:LippmannSchwinger} still holds, and we can follow the formal steps in exactly the same way as above, first projecting into position space
\beq
\braket{\mathbf{x}|\psi^\pm} &=&  \braket{\mathbf{x}|\phi} + \int d^3 x'\braket{\mathbf{x}|\frac{1}{E- H_0 \pm i \epsilon}|\mathbf{x}'}\braket{\mathbf{x}'|V|\psi^\pm} \nonumber \\
&=&  \braket{\mathbf{x}|\phi} + \int d^3 x'\, d^3 k\, d^3 k'\braket{\mathbf{x}|\mathbf{k}}\braket{\mathbf{k}|\frac{1}{E- H_0 \pm i \epsilon}|\mathbf{k'}}\braket{\mathbf{k}|\mathbf{x}'}\braket{\mathbf{x}'|V|\psi^\pm} \nonumber \\
&=&  \braket{\mathbf{x}|\phi} + \int_{-L/2}^{L/2} d^3 x'\, d^3 k \,\frac{e^{i(\mathbf{x}-\mathbf{x'})\cdot \mathbf{k}}}{(2\pi)^3}\braket{\mathbf{k}|\frac{1}{E+\frac{1}{2m}\nabla^2\pm i \epsilon}|\mathbf{k'}}\braket{\mathbf{x}'|V|\psi^\pm} \nonumber \\
&=&  \braket{\mathbf{x}|\phi} + \int_{-L/2}^{L/2} d^3 x'\, d^3 k\,\frac{e^{i(\mathbf{x}-\mathbf{x'})\cdot \mathbf{k}}}{(2\pi)^3}\frac{1}{E- \frac{k^2}{2m} \pm i \epsilon}\braket{\mathbf{x}'|V|\psi^\pm}  \nonumber \\
&=&\braket{\mathbf{x}|\phi} - m \frac{e^{i k r}}{2\pi r}\int_{-L/2}^{L/2} d^3 x'\, e^{-i \mathbf{x'}\cdot \mathbf{k}'}V(\mathbf{x}')\psi^\pm(\mathbf{x'})
\label{eqn:FiniteLocalization}
\eeq
where in the last line we used the same assumptions as in the standard case that the detector is very far away from the scattering centers, {\textit i.e} $|\mathbf{x}| \gg R$. 

Eq.~\eqref{eqn:FiniteLocalization} is formally then very similar to the case where $H_0$ is simply the free Hamiltonian, except that the effect of $V_W$ is to restrict the Fourier integral in the position domain. In anaology with the free scattering case, we can define a scattering amplitude\,,
\beq
f^{(1)}_\text{loc.}(\mathbf{k}, \mathbf{k}') = -\frac{m}{2\pi} \int_{-L/2}^{L/2} d^3x' \, e^{i  \mathbf{x}'\cdot \left(\mathbf{k}-\mathbf{k}'\right)}V(\mathbf{x}') \,.
\eeq
Note that, for $L> R$, {\textit i.e.} the wavefunction extent is larger than the size of scattering center, $f^{(1)}_\text{loc.}(\mathbf{k}, \mathbf{k}')=f^{(1)}_\text{free}(\mathbf{k}, \mathbf{k}')$, as we would expect. However, if $L < R$, then only the part of the scattering center than overlaps with the scattered particle's wavefunction contributes to the scattering, that is, $N_S$ in the above is replaced by $ N_S L^3/R^3$. 
Explicitly, when $q L \ll 1$, the total cross section is
\beq
\left(\frac{d\sigma}{d\Omega}\right)_\text{Coh} = \left(N_S \frac{L^3}{R^3}\right)^2\, \left(\frac{d\sigma_1}{d\Omega}\right)_\text{free}
\eeq
and when $q L \gtrsim 1$, the total cross section is
\beq
\left(\frac{d\sigma}{d\Omega}\right)_\text{Incoh} = \left(N_S \frac{L^3}{R^3}\right)\, \left(\frac{d\sigma_1}{d\Omega}\right)_\text{free} 
\eeq
where $\left(d\sigma_1/d\Omega\right)_\text{free}$ is the single scatterer cross section, calculated in the place wave approximation. 

This result, that only the scattering centers that have a non-zero overlap with the scattered particle's wavefunction contribute to the scattering cross section, is the basis of Eqs.~\eqref{eq:fermionshortdedx} and \eqref{eq:fermionshortdedxCoh}.

\bibliography{bibliography.bib}

\end{document}